\documentclass[aps,prd,onecolumn,showpacs,eqsecnum,nofootinbib]{revtex4}
\usepackage{epsfig}
\usepackage{graphicx}
\usepackage{dcolumn}
\usepackage{amsmath}
\usepackage{enumerate}

\def\beq{\begin{equation}}
\def\eeq{\end{equation}}

 \def\gtap{\mathrel{ \rlap{\raise 0.511ex \hbox{$>$}}{\lower 0.511ex
   \hbox{$\sim$}}}} 
\def\ltap{\mathrel{ \rlap{\raise 0.511ex
    \hbox{$<$}}{\lower 0.511ex \hbox{$\sim$}}}} 
\newcommand{\bea}{\begin{eqnarray}} \newcommand{\eea}{\end{eqnarray}}

\newcommand{\tetaot}{\mbox{$\theta_{13}$}}

\newcommand{\nova}{\mbox{NO$\nu$A}}
\newcommand{\novasp}{\mbox{NO$\nu$A} }

\begin{document}

\vskip-6pt \hfill {IPPP/07/60} \\
\vskip-6pt \hfill {DCPT/07/120} \\
\vskip-6pt \hfill {FERMILAB-PUB-07-480-E} \\
\vskip-6pt \hfill {Roma-TH-1459} \\
\title{A Neutrino Factory for both Large and Small $\theta_{13}$}

\author{Alan Bross}
\author{Malcolm Ellis}
\author{Steve Geer}
\email{sgeer@fnal.gov}
\affiliation{Fermi National Accelerator Laboratory, Batavia, IL 60510-0500, USA}
\author{Olga Mena} 
\email{omena@fnal.gov}
\affiliation{INFN - Sez. di Roma,Dipartimento di Fisica, Universit\'a di Roma "La Sapienza", P.le A. Moro, 5, I-00185 Roma, Italy}
\author{and Silvia Pascoli}
\email{silvia.pascoli@durham.ac.uk}
\affiliation{IPPP, Department of Physics, Durham University, Durham DH1 3LE, United Kingdom}

\begin{abstract}
An analysis of the neutrino oscillation physics capability of a low energy Neutrino Factory is presented, including a first simulation of the detector efficiency and event energy threshold. The sensitivity of the physics reach to the presence of backgrounds is also studied. We consider a representative baseline of $1480$ km, we use muons with 4.12 GeV energy and we exploit a very conservative estimate of the energy resolution of the detector. Our analysis suggests an impressive physics reach for this setup, which can eliminate degenerate solutions, for both large and small values of the mixing angle $\theta_{13}$, and can determine leptonic CP violation and the neutrino mass hierarchy with extraordinary sensitivity.

\end{abstract}


\maketitle

\section{Introduction}

In recent years compelling evidence for neutrino oscillations has been found in experiments with atmospheric~\cite{SKatm}, solar~\cite{sol,SKsolar,SNO1,SNO2,SNO3,SNOsalt}, reactor~\cite{KamLAND} and long-baseline accelerator neutrinos~\cite{K2K,MINOS}. Two mass squared differences, $\Delta m^2_{ji} \equiv m_j^2 - m_i^2$, have been measured with good accuracy, their present best fit values being $|\Delta m^2_{31} |= 2.5 \times 10^{-3} \ \mathrm{eV}^2$ and $|\Delta m^2_{21} |= 8.0 \times 10^{-5} \ \mathrm{eV}^2$. 
In addition, explaining the experimental data in terms of neutrino oscillations requires two large mixing angles in the leptonic mixing matrix $U$. Their best fit values are:
 $\sin^2 \theta_{12} = 0.30$ and $\sin^2 2 \theta_{23} =1$, see Refs.~\cite{thomas,newfit,concha}.  
 Despite the remarkable recent progress in our understanding of neutrino physics, fundamental questions remain unanswered. It is crucial to establish the nature of neutrinos - whether they are Dirac or Majorana particles, the neutrino mass ordering, the absolute neutrino mass scale, the value of the unknown mixing angle $\theta_{13}$, the presence or absence of CP violation in the leptonic sector, and the precise values of the already known oscillation parameters. This information will help shed light on the physics beyond the Standard Model responsible for neutrino masses and for the leptonic mixing structure.
 
In order to achieve these goals, very sensitive neutrino experiments will be required. In particular, long baseline oscillation experiments are expected to play an important role in providing precision measurements of the neutrino oscillation parameters, the CP-violating phase $\delta$, and a determination of the neutrino mass ordering.

Neutrino Factories~\cite{geer}, in which a neutrino beam is generated from muons decaying within the straight sections of a storage ring, have been studied extensively in the past, and have been shown to be sensitive tools for studying neutrino oscillation physics~\cite{geer,nf1,nf2,nf4,nf5,nf6,silver,study1-physics,nf8,study2,yo,nf9,GeerMenaPascoli,hubernew}. In a Neutrino Factory far detector, the experimental signature for the so called \emph{golden channel}~\cite{nf5} is the presence of a \emph{wrong-sign} muon~\cite{geer,nf1}, i.e. a muon with opposite sign to the muons stored in the neutrino factory. Wrong-sign muons result from $\nu_{e} \rightarrow \nu_{\mu}$ oscillations, and can be used to measure the unknown mixing angle $\theta_{13}$, determine the neutrino mass hierarchy, and search for CP violation in the neutrino sector. This physics program requires the detection of charged current (CC) muon-neutrino interactions, and the measurement of the sign of the produced muon. If the interacting neutrinos have energies of more than a few GeV, standard neutrino detector technology, based on large magnetized sampling calorimeters, can be used to measure wrong-sign muons with high efficiency and very low backgrounds. This has been shown to work for Neutrino Factories with energies of about 20 GeV or greater~\cite{nf5,ISS-Detector Report,study1-physics}. 

Lower energy Neutrino Factories~\cite{GeerMenaPascoli}, which store muons with energies $<10$~GeV, require a detector technology that can detect lower energy muons. Recently ideas have emerged for a Neutrino Factory detector based on a fully active calorimeter within a potentially affordable large volume magnet. These ideas encourage consideration of low energy Neutrino Factories. Initial studies~\cite{GeerMenaPascoli}, based on a first guess for the performance of a low energy Neutrino Factory detector, suggested that a Neutrino Factory with an energy of about 4 GeV would enable very precise measurements of the neutrino mixing parameters. 
In the present paper, we consider in more detail the expected performance of a low energy Neutrino Factory detector, and update and extend our previous studies to include a more realistic detector model and a more comprehensive study of systematic effects. 
In particular we exploit the low energy threshold of the detector and
make a very conservative estimate for its energy resolution which, together with the broad spectrum of the neutrino factory beam, facilitates the elimination of degeneracies~\cite{FL96,MN01,BMWdeg,deg}. It is well known that even a very precise measurement of the appearance probability for neutrinos and antineutrinos at a fixed $L/E$ allows different solutions for $(\theta_{13}, \mathrm{sign} (\Delta m^2_{13}), \delta)$, weakening severely the sensitivity to these parameters. Many strategies have been advocated to resolve this issue which in general involve another detector~\cite{MN97,silver,BMW02off,SN1,twodetect,SN2,T2kk} or the combination with another experiment~\cite{otherexp1,BMW02,HLW02,MNP03,otherexp,mp2,HMS05,Choubey,yo,nf9,huber2,lastmine}.
Using the energy dependence of the signal in the low energy Neutrino Factory, 
we find that a 4 GeV Neutrino Factory can unambiguously determine all of the neutrino oscillation parameters with good precision provided $\sin^2 2\theta_{13}>$ few $\cdot 10^{-3}$. Hence a low energy Neutrino Factory would be a precision tool for both large and small $\theta_{13}$.
  
In Section II we describe the design for the low-threshold detector and its performance. In Section III, we discuss in detail the physics reach of the proposed setup. We first consider the disappearance $\nu_\mu$ signal in order to determine precisely the value of the atmospheric mass squared difference and, possibly, the type of hierarchy even for $\theta_{13}=0$.
Then, we consider the appearance signals $\nu_e \rightarrow \nu_\mu$ and $\bar{\nu}_e \rightarrow \bar{\nu}_\mu$, which depend on $\theta_{13}$, $\delta$ and the type of neutrino mass ordering. 
We perform a detailed numerical simulation and discuss the sensitivity of the low-energy Neutrino Factory to these parameters.
In Section IV, we draw our conclusions.

\section{Detector design and performance}

A totally active scintillator detector (TASD) has been proposed for a Neutrino Factory, and results from a first study of its expected performance are  described in the recent International Scoping Study Report~\cite{ISS-Detector Report}. Using a TASD for neutrino physics is not new.  Examples are KamLAND~\cite{KamLAND}, which has been operating for several years, and the proposed NO$\nu$A detector~\cite{Nova}, which is a $15-18$~Kton liquid scintillator detector that will operate off-axis to the NuMI beam line~\cite{Numi} at Fermilab. Note that, unlike KamLAND or NO$\nu$A, the TASD we are investigating for the low energy Neutrino Factory is magnetized and has a segmentation that is approximately 10 times that of \nova. Magnetization of such a large volume ($>30,000$~m$^3$) is the main technical challenge in designing a TASD for a Neutrino Factory, although R\&D to reduce the detector cost (driven in part by the large channel count, $7.5 \times 10^6$) is also needed. 

The Neutrino Factory TASD we are considering consists of long plastic scintillator bars with a triangular cross-section arranged in planes which
 make x and y measurements (we plan to also consider an x-u-v readout scheme).  Optimization of the cell cross section still needs further study since a true triangular cross section results in tracking anomalies at the corners of the triangle.  The scintillator bars have a length of $15$~m and
the triangular cross-section has a base of $3$~cm and a height of $1.5$~cm. We have considered a design using liquid as in \nova, but, compared to \nova, the cell size is small (\novasp uses a $4\times 6$~cm$^2$ cell) and the non-active component due to the PVC extrusions that hold the liquid becomes quite large (in \nova, the scintillator is approximately $70\%$ of the detector mass). Our design is
an extrapolation of the MINER$\nu$A experiment~\cite{minerva_www} which in turn was an extrapolation of the D0
preshower detectors~\cite{D0}.  We are considering a detector mass of approximately $35$~Kton (dimensions $15 \times 15 \times 150$~m).
We believe that an air-core solenoid can produce the field required ($0.5$~Tesla) to do the physics.

\begin{figure}[h]
\begin{center}
\includegraphics[width=6.5in]{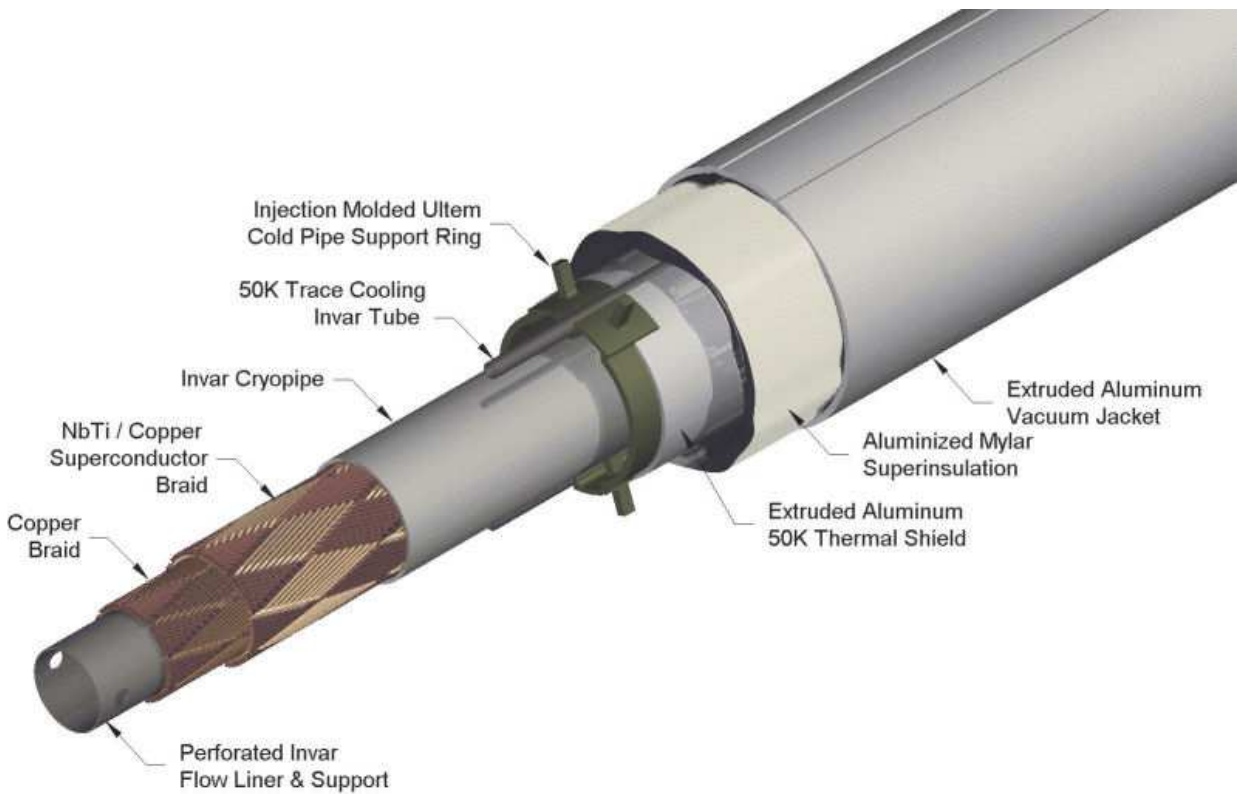}
\end{center}
\caption[]{\textit{Diagram of Superconducting Transmission Line design.}}
\label{fig:STL}
\end{figure}

As was mentioned above, magnetizing the large detector volume presents the main technical challenge for a Neutrino Factory TASD.  Conventional
room temperature magnets are ruled out due to their prohibitive power consumption, and conventional superconducting magnets are believed to be too expensive, due to the cost of the enormous cryostats needed in a conventional superconducting magnet design. In order to eliminate 
the cryostat, we have investigated a concept based on the  superconducting transmission line (STL) that was 
developed for the Very Large Hadron Collider superferric magnets~\cite{VLHC}. The solenoid windings now consist of this superconducting cable which is confined 
in its own cryostat (Fig.~\ref{fig:STL}).  Each solenoid ($10$ required for the full detector) consists of $150$ turns and requires $7500$ m of cable. There is 
no large vacuum vessel and thus no large vacuum loads which make the cryostats for large conventional superconducting magnets very expensive. 

\begin{figure}[h]
\begin{center}
\includegraphics[width=4in]{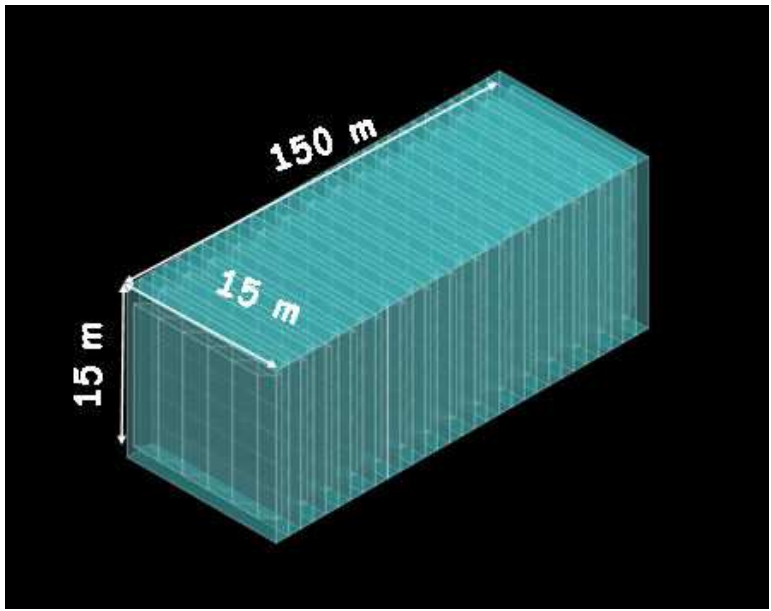}
\end{center}
\caption[]{\textit{Schematic of Totally Active Scintillator Detector.}}
\label{fig:tasd}
\end{figure}

The Neutrino Factory TASD response has been simulated with GEANT4 version 8.1 (Fig.~\ref{fig:tasd}).  The GEANT4 model of the detector included each of the individual scintillator bars, but did not include edge effects on light collection, or the effects of a central wavelength shifting fiber. A uniform 0.5 Tesla magnetic field was simulated.
 
Samples of isolated muons in the range of momentum between $100$~MeV$/c$ and $15$~GeV$/c$ were simulated to allow the determination of the momentum resolution and charge identification capabilities. The NUANCE~\cite{Nuance} event generator was also used to simulate 1 million $\nu_e$ and 1 million $\nu_\mu$ interactions. Events were generated in $50$ mono-energetic neutrino energy bins between $100$~MeV and $5$~GeV. The results that follow only have one thousand events processed through the GEANT4 simulation and reconstruction.
 
The detector response was simulated assuming a light yield consistent with MINER$\nu$A measurements and current photo detector performance~\cite{Nova}. In addition, a 2 photo-electron energy resolution was added through Gaussian smearing to ensure that the energy resolution used in the following physics analysis would be a worst-case estimate. Since  a complete pattern recognition algorithm was beyond the scope of our study, for our analysis the Monte Carlo information was used to aid in pattern recognition. All digitised hits from a given simulated particle where the reconstructed signal was above 0.5 photo electrons were collected. When using the isolated particles, hits in neighboring x and y planes were used to determine the 3 dimensional position of the particle. The position resolution was found to be approximately $4.5$~mm RMS with a central Gaussian of width $2.5$~mm~\footnote{At this stage, the simulation does not take into account light collection inefficiencies in the corners of the base of the triangle.}. These space points were then passed to the RecPack Kalman track fitting package~\cite{recpack}. 
 
\begin{figure}[h]
\begin{center}
\includegraphics[width=4in]{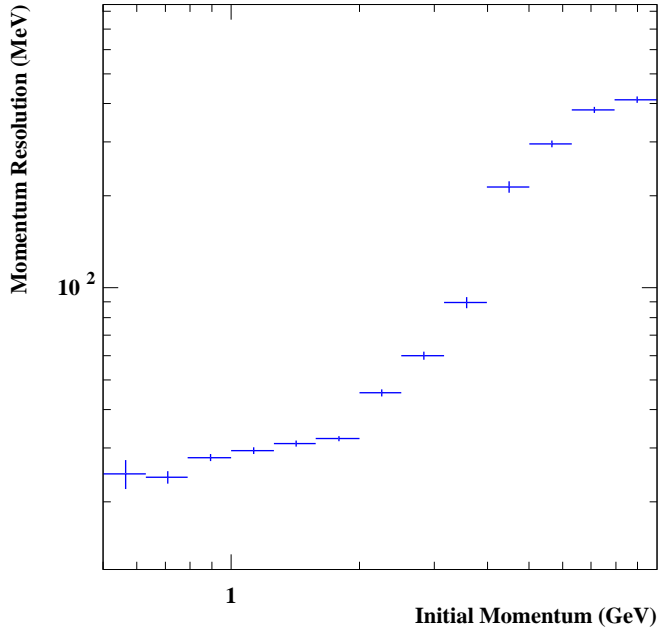}
\end{center}
\caption[]{\textit{Momentum resolution as a function of the muon momentum.}}
\label{fig:tasd_mom}
\end{figure}

\begin{figure}[h]
\begin{center}
\begin{tabular}{ll}
\includegraphics[width=3in]{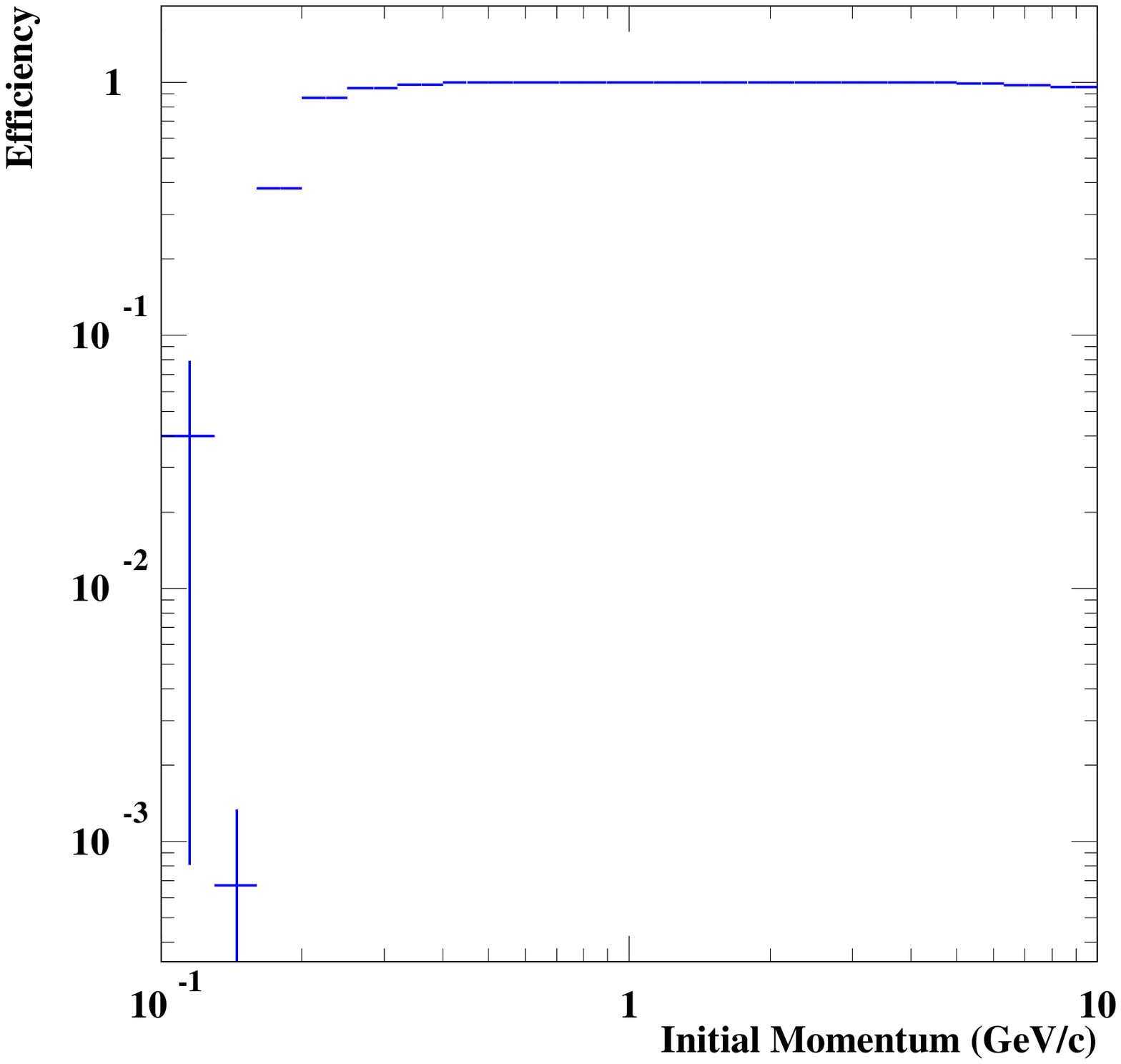}&\hskip 0.cm
\includegraphics[width=3in]{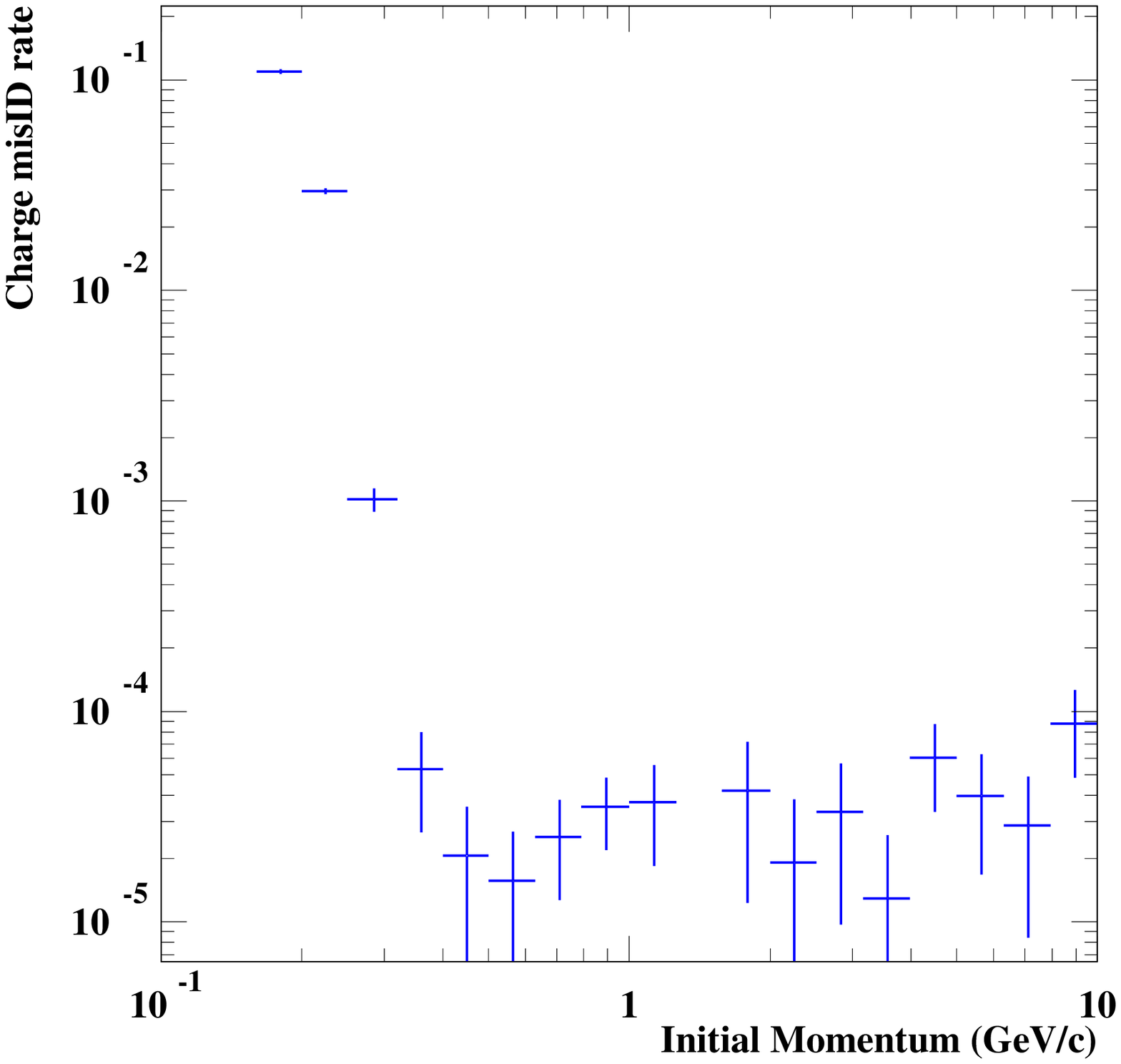}\\
\hskip 4.truecm
{\small (a)}            &
\hskip 4.truecm
{\small (b)}       \\  
\end{tabular}
\end{center}
\caption{\textit{(a) Efficiency for reconstructing positive muons. (b)
Muon charge mis-identification rate as a function of the initial muon momentum.}}
\label{fig:Track}
\end{figure}

For each collection of points, the track fit was performed with an assumed positive and negative charge. The momentum resolution and charge misidentification rates were determined by studying the fitted track in each case which had the better $\chi^2$ per degree of freedom. Figure~\ref{fig:tasd_mom} shows the momentum resolution as a function of muon momentum. The tracker achieves a resolution of better than $10\%$ over the momentum range studied. Figure~\ref{fig:Track}(a) shows the efficiency for reconstructing positive muons as a function of the initial muon momentum. The detector becomes fully efficient above $400$~MeV.
 
The charge mis-identification rate was determined by counting the rate at which the track fit with the incorrect charge had a better $\chi^2$ per degree of freedom than that with the correct charge. Figure~\ref{fig:Track}(b) shows the charge mis-identification rate as a function of the initial muon momentum.
 
The neutrino interactions were also reconstructed using the aid of the Monte Carlo information for pattern recognition. In an attempt to produce some of the effects of a real pattern recognition algorithm on the detector performance, only every fourth hit was collected for track fitting. Tracks were only fit if $10$ such hits were found from a given particle. The Monte Carlo positions were smeared (Gaussian smearing using the $4.5$~mm RMS determined previously) and passed to the Kalman track fit. The reconstruction returned:
\begin{itemize}
\item The total momentum vector of all fitted tracks,
\item The momentum vector of the muon (muon ID from MC truth),
\item The reconstructed and truth energy sum of all the hits that were not in a particle that was fitted, and 
\item The reconstructed energy sum of all hits in the event.
\end{itemize}

\begin{figure}[h]
\begin{center}
\begin{tabular}{ll}
\includegraphics[width=3in]{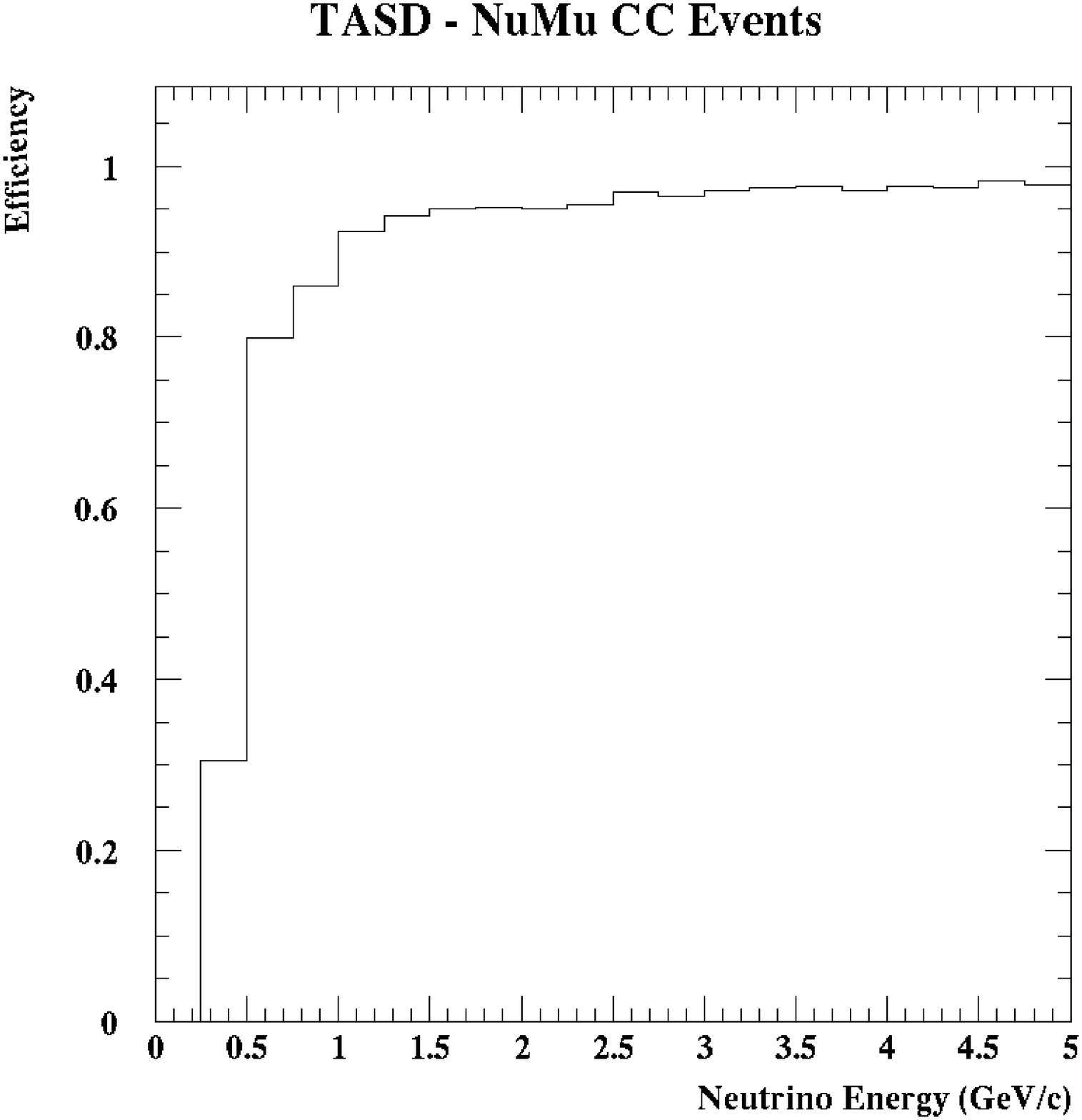}&\hskip 0.cm
\includegraphics[width=3in]{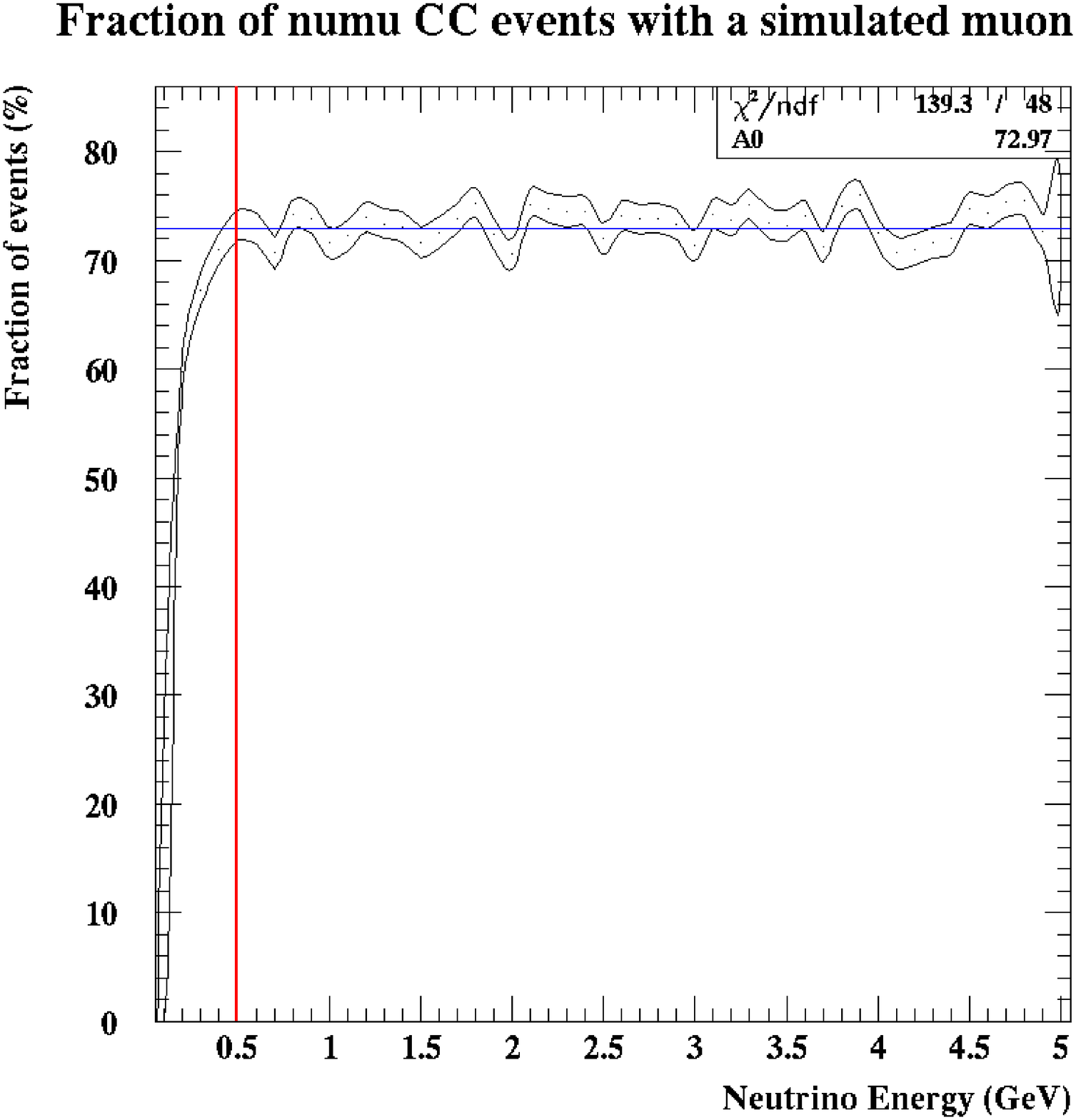}\\
\hskip 4.truecm
{\small (a)}            &
\hskip 4.truecm
{\small (b)}       \\  
\end{tabular}
\end{center}
\caption{\textit{(a) Reconstruction efficiency of NuMu CC events as a function of neutrino interaction energy. (b) Fraction of NuMu CC events with a reconstructed muon.}}
\label{fig:NuMuCC}
\end{figure}

The $\nu_{\mu}$ CC event reconstruction efficiency as a function of neutrino energy is shown in Fig.~\ref{fig:NuMuCC}(a).  The fraction of
$\nu_{\mu}$ CC events with a reconstructed muon is shown in Fig.~\ref{fig:NuMuCC}(b).  In this figure the bands represent the limits of the 
statistical errors for this analysis.

Based on these initial Neutrino Factory TASD studies, in our phenomenological analysis we assume the detector has an effective threshold for measuring muon neutrino CC events at $E_\nu = 500$~MeV, above which it has an energy independent efficiency of $73\%$. The $73\%$ efficiency is primarily driven by the neutrino interaction kinematics, not by the detector tracking efficiency.  No charge-ID criterion is applied here. The charge misidentification rate information is used as input into the effect of backgrounds on the analysis.

We note that, to fully understand the backgrounds in the TASD requires a simulation that includes neutrino interactions and a full event reconstruction.  Although this is beyond the scope of the present study, a consideration of backgrounds in the well studied Magnetized Fe-Scintillator detector proposed for the high-energy Neutrino Factory~\cite{nf5} and The International Scoping Study for a Neutrino Factory~\cite{ISS-Detector Report} motivates the $10^{-3}$ background (contamination) assumption used in this paper for the TASD. Before kinematic cuts, the main backgrounds for the Fe-Scintillator detector are muon charge mis-ID, charm decay, pion and kaon decay, and are all of comparable order: $1-5\times 10^{-4}$.  For the TASD at a low energy Neutrino Factory the muon charge mis-ID rate (Fig.~\ref{fig:Track}(b)) and the charm decay background is suppressed (at the level $4-8\times 10^{-5}$) due to the low energy beam.  Pion and Kaon decay in flight become the main background concerns at the $1-5\times 10^{-4}$ level.  A figure of merit for comparing the TASD to a conventional Magnetized Fe-Scintillator detector is the ratio of their respective particle decay lengths to interaction lengths.  For TASD the ratio is about $1$; for the Magnetized Fe-Scintillator detector it is approximately $8$.  So naively we can conclude that the decay background in TASD will be 10 times worse than in the conventional detector ignoring any kinematic or topological cuts.  However the TASD will have vastly superior kink detection to identify decay-in-flight.  For example we will typically have $40$ hits on the pion track before decay.  In addition TASD will have continuous $dE/dx$ measurements along the track and better overall energy resolution.  We believe that these properties will allow us to control backgrounds to the $10^{-3}$ level or better.

\section{Physics reach of the low energy Neutrino Factory}

We have previously mentioned that, by exploiting the energy dependence of the signal, it is possible to extract from the measurements the correct values of $\theta_{13}$ and $\delta$, and eliminate the additional solutions arising from discrete ambiguities. In the present study, we include the detector simulation results described in the previous section, which suggests a lower energy threshold (500~MeV) than previously assumed~\cite{GeerMenaPascoli}, and an energy resolution $dE/E=30\%$~\footnote{We have assumed a very conservative $dE/E=30\%$ because at this time the
simulation work has not yet produced a number for the TASD.  Based on \novasp
results, we expect the TASD $dE/E$ to be better than $6\%$ at 2~GeV.}. Above threshold, the detector efficiency for muon neutrino CC events is taken to be 73\%.

In the following we consider the representative baseline $L=1480$~km, which corresponds to the distance from Fermilab to the Henderson mine. However, we believe that the TASD will not require operation deep underground in order to remove backgrounds. Results are similar for other baselines in the 1200--1500~km range. The results are presented for the high-statistics scenario described in~\cite{GeerMenaPascoli} as well as for a more aggressive scenario which improves the statistics of the old high-statistics scenario by a factor of three, to quantify the benefits of increased detector sizes and/or stored-muon luminosities. The high-statistics scenario corresponds to $1 \times 10^{23}$~Kton-decays (10 years of data taking, with $5 \times 10^{20}$ useful muon decays of each sign per year, and a detector fiducial mass times efficiency of 20~Kt). The more aggressive scenario corresponds to $3 \times 10^{23}$~Kton-decays (which could correspond, for instance, to 10 years of data taking, with $1 \times 10^{21}$ useful muon decays of each sign per year, and a detector fiducial mass times efficiency of 30~Kt).

Table~\ref{tab:tab1} shows the number of CC muon events expected in the two scenarios explored here for, respectively, the positive and negative muons stored in the Neutrino Factory. Notice that, in the absence of oscillations, there would be a few times $10^4$~$\nu_e$ CC interactions, which would allow a search for $\nu_e \rightarrow \nu_\mu$ oscillations with probabilities below $10^{-4}$.
\begin{table}[thb]
\centering
\begin{tabular}{||c|c||c|c||c|c||}
\hline \hline
\multicolumn{2}{||c||}{$E_{\mu^{\mp}}=$} & 
\multicolumn{2}{c||} {$\mu^{+}$} &  
\multicolumn{2}{c||}{$\mu^{-}$}\\
\cline{3-6}
\multicolumn{2}{||c||}{$4.12$ GeV}& $N_{\bar{\nu}_\mu}/10^3$ & $N_{\nu_e}/10^3$  & 
$N_{\nu_\mu}/10^3$ &$N_{\bar{\nu}_e}/10^3$ \\  
\hline\hline
statistics& $1$& $13$ & $22$ & $25$ & $11$\\
\cline{2-6}
$(10^{23})$ Kt-decays
   & $3$& $39$ & $66$ & $77$ & $34$\\
\hline\hline
\end{tabular}
\caption{\it{ 
Neutrino and antineutrino charged currents interaction rates 
for L = 1480~km, for the $10^{23}$~Kton-decays and the $3 \times 10^{23}$~Kton-decays-statistics scenarios.}} 
\label{tab:tab1}
\end{table}

All numerical results reported in the next subsections have been obtained with the exact formulae for the oscillation probabilities. Unless specified otherwise, we take the following central values for the remaining oscillation parameters: $\sin^{2}\theta_{12}=0.29$, $\Delta m^2_{21} =  8 \times 10^{-5}$ eV$^2$,  $|\Delta m^2_{31}| = 2.5 \times 10^{-3}$ eV$^2$ and  $\theta_{23}=40^\circ$. We show in Tables~\ref{tab:tab2} and \ref{tab:tab3}, for two representative values of $\theta_{13}=1^\circ$ and $8^\circ$, and the CP phase $\delta=0^\circ, 90^\circ, 180^\circ$ and $270^\circ$, the number of \emph{wrong-sign} muon events in the two scenarios explored here, for, respectively, the positive and negative muons stored in the Neutrino Factory, for normal (inverted) hierarchy.
\begin{table}[bht]
\centering
\begin{tabular}{||c|c||c||c||}
\hline\hline
statistics (Kt-decays) & $\delta(^{o})$ & $\mu^{+}$ stored (wrong-sign: $\mu^-$)  &$\mu^{-}$ stored (wrong-sign: $\mu^{+}$) \\
\hline\hline
     & 0 & 880 (340) & 180 (520)\\
\cline{2-4}
$1 \times 10^{23}$     
     & 90 & 1230 (505) & 90 (330)\\
\cline{2-4}
     & 180 & 1000 (340) & 170 (440)\\
\cline{2-4}
     & 270 & 645 (175) & 260 (625) \\
\hline\hline  
       & 0 &2640 (1020)& 540 (1550)\\ 
\cline{2-4}
$3 \times 10^{23}$     
     & 90 & 3700 (1520) &270 (990)\\
\cline{2-4}
     & 180 &2990 (1020) & 510 (1310)\\
\cline{2-4}
     & 270 &1930 (520) &780 (1870) \\
\hline\hline 
\end{tabular}
\caption{\it 
{Wrong sign muon event rates for normal (inverted) hierarchy, assuming $\nu_e\to \nu_\mu$ ($\bar{\nu}_e \to \bar{\nu}_\mu$)
oscillations in a 20~Kt fiducial volume detector, for a L = 1480~km baseline. 
We assume here $\theta_{13}=8^{o}$, i.e. $\sin^2 2\theta_{13}\simeq 0.076$. 
We present the results for several possible values of the CP-violating phase $\delta$  for both the two scenarios.}}
\label{tab:tab2}
\end{table} 
\begin{table}[bht]
\centering
\begin{tabular}{||c|c||c||c||}
\hline\hline
statistics (Kt-decays)& $\delta(^{o})$ & $\mu^{+}$ stored (wrong-sign: $\mu^-$)  &$\mu^{-}$ stored (wrong-sign: $\mu^{+}$) \\
\hline\hline
     & 0 & 54 (50) &  27 (37) \\
\cline{2-4}
$1 \times 10^{23}$     
     & 90 & 100 (70) & 13 (10)\\
\cline{2-4}
     & 180 & 67 (50) & 70 (25)\\
\cline{2-4}
     & 270 & 22 (30) & 37 (50) \\
\hline\hline  
       & 0 &160 (150)& 80 (110)\\ 
\cline{2-4}
$3 \times 10^{23}$     
     & 90 & 300 (210)&40 (30)\\
\cline{2-4}
     & 180 &200 (150)& 230 (250)\\
\cline{2-4}
     & 270 &65 (90) &110 (150)\\
\hline\hline 
\end{tabular}
\caption{\it 
{As Table~\protect\ref{tab:tab2} but for  $\theta_{13}=1^{o}$, i.e. $\sin^2 2\theta_{13}\simeq 0.001$. 
}}
\label{tab:tab3}
\end{table}

For our analysis, we use the following $\chi^{2}$ definition
\begin{equation}
\chi^2 = \sum_{i,j} \sum_{p,p'} \; (n_{i,p} - N_{i,p}) C_{i,p:,j,p'}^{-1} (n_{j,p'} - N_{j,p'})\,,
\end{equation}
 where $N_{i,\pm}$ is the predicted number of muons for
 a certain oscillation hypothesis, $n_{i,p}$ are the simulated ``data'' from a Gaussian or Poisson smearing and $C$ is the $2 N_{bin} \times 2 N_{bin}$ covariance matrix given by:
\begin{equation}
C_{i,p:,j,p'}^{-1}\equiv \delta_{ij}\delta_{pp'}(\delta n_{i,p})^2 
\end{equation}
where $(\delta n_{i,p}) = \sqrt{n_{i,p} + (f_{sys}\cdot n_{i,p})^2}$
contains both statistical and a $2\%$ overall systematic error ($f_{sys}=0.02$).

\subsection{Exploring the disappearance channel}

Consider first the disappearance channels, already considered in the context of Neutrino Factories~\cite{nf4,dis} and carefully explored in Ref.~\cite{Stef}.
In Ref.~\cite{GeerMenaPascoli} it was shown that, with its high statistics and good energy resolution, a low energy neutrino factory can be used to precisely determine the atmospheric neutrino oscillation parameters, $\theta_{23}$ and $\Delta m^2_{31}$. In particular, for an exposure of $3 \times 10^{22}$~Kton-decays for each muon sign, and allowing for a $2\%$ systematic uncertainty, it was shown that: (i) 
Maximal mixing in the 23-sector could be excluded at $99\%$ CL if $\sin^2 \theta_{23}<0.48$ ($\theta_{23}<43.8^\circ$), independently of the value of $\theta_{13}$, and (ii) For a large value of $\theta_{13}$, i.e. $\theta_{13}>8^\circ$, the $\theta_{23}$-octant degeneracy would be resolved at the $99\%$ CL for $\sin^2 \theta_{23}<0.44$ ($\theta_{23} < 41.5^\circ$). 
In our present study, the good energy resolution of the TASD provides sensitivity to the oscillatory pattern of the disappearance signal that is comparable to, and somewhat better than, we previously assumed. 

\begin{figure}[h]
\begin{center}
\begin{tabular}{ll}
\includegraphics[width=3in]{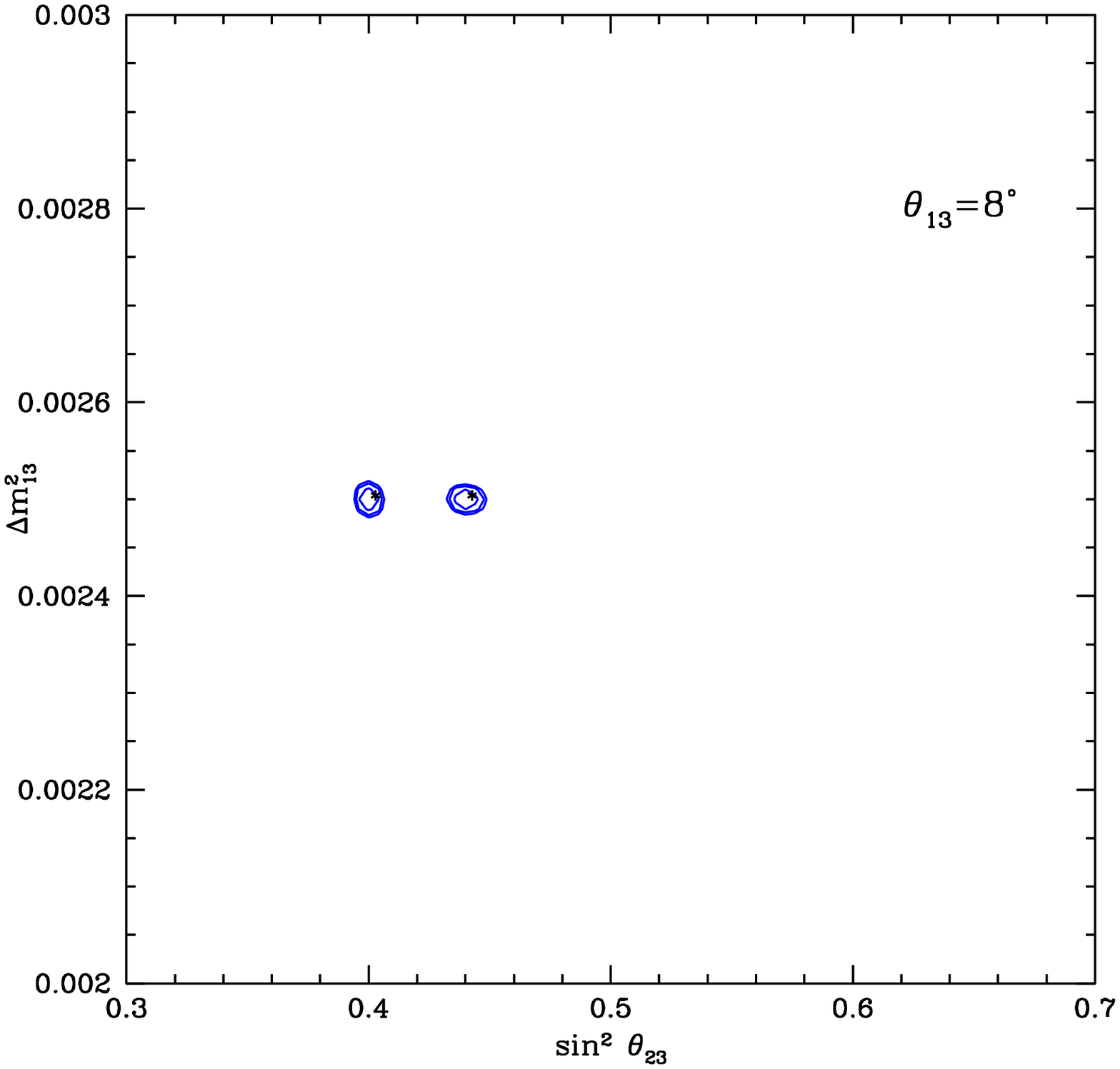}&\hskip 0.cm
\includegraphics[width=3in]{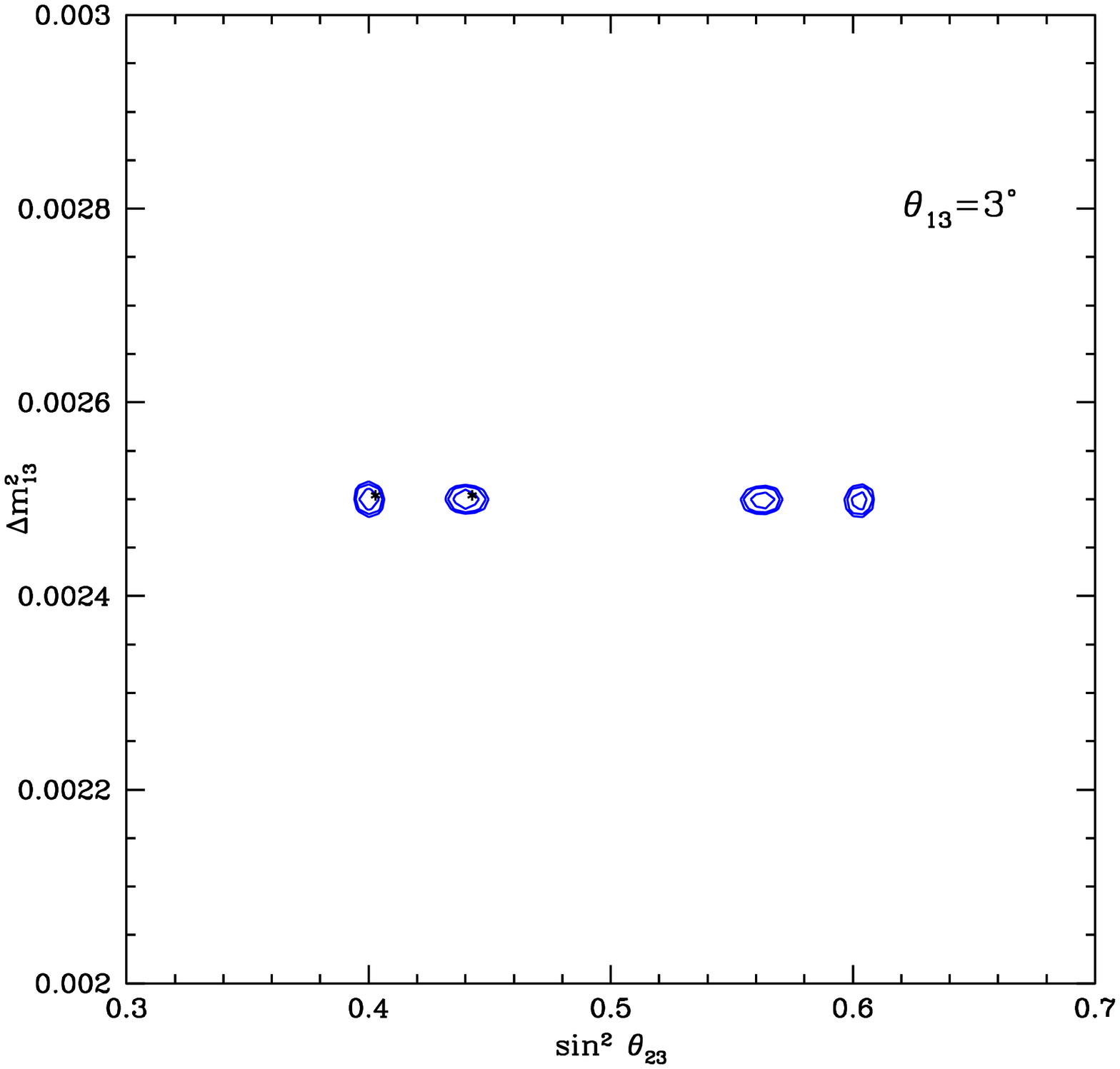}\\
\end{tabular}
\end{center}
\caption{\textit{$68\%$, $90\%$ and $95\%$ (2 d.o.f)  CL contours resulting from the fits at $L = 1480$ km assuming two central values for $\sin^2 \theta_{23}=0.4$ and $0.44$ and  $\Delta m^2_{31}=2.5 \times 10^{-3}$ eV~$^2$. In the left (right) panel, $\theta_{13}=8^\circ$ ($3^\circ$). The statistics considered for both simulations corresponds to $1\times 10^{23}$~Kton-decays.  Only disappearance data have been used to perform these plots.} }  \label{fig:dis}
\end{figure}

In Fig.~\ref{fig:dis} we show the $68\%$, $90\%$ and $95\%$~CL contours (for 2 d.o.f) resulting from the fits to the measured energy dependent $\nu_\mu$ and $\overline{\nu}_{\mu}$ CC rates at $L = 1480$ km. Results correspond to $1 \times 10^{23}$~Kton-decays, and are shown for $\Delta m^2_{31}=2.5 \times 10^{-3}$ eV~$^2$ and two simulated values of $\sin^2 \theta_{23}$ (= $0.4$ and $0.44$). For $\theta_{13}=0$, $P_{\nu_\mu \to \nu_\mu} (\theta_{23})=  P_{\nu_\mu \to \nu_\mu} (\pi/2 -\theta_{23})$, i.e. the disappearance channel is symmetric under $\theta_{23}\to \pi/2-\theta_{23}$. However, when a rather large non-vanishing value of $\theta_{13}$ is switched on, a $\theta_{23}$ asymmetry appears in the $P_{\nu_\mu \to \nu_\mu}$. Notice that the asymmetry grows with increasing $\theta_{13}$ and the four-fold degeneracy in the atmospheric neutrino parameters is resolved more easily. We conclude that, using only the $\nu_\mu$-disappearance data, the uncertainty on $\Delta m^2_{31}$ could be reduced down to the $1\%-2\%$ level. In principle, the $\nu_e$ disappearance channel could also be used, which is sensitive to $\theta_{13}$ and matter effects. However, charge discrimination for electrons has not yet been adequately studied to determine the relevant TASD performance parameters.

The extremely good determination of the atmospheric mass squared difference opens the possibility 
to determine the mass hierarchy by exploiting the effects of the solar mass squared difference on the $\nu_\mu$ disappearance probability, even for negligible values of $\theta_{13}$.
This strategy was studied in detail in Refs.~\cite{deGouvea:2005hk,deGouvea:2005mi,Minakata:2006gq}.  The vacuum $\nu_\mu \to \nu_\mu$ oscillation probability is given by
\begin{equation}
P(\nu_\mu \to \nu_\mu) = 1 - 4 | U_{\mu 1}|^2 | U_{\mu 2}|^2 \sin^2 \frac{\Delta m^2_{12} L}{4E} - 4 | U_{\mu 1}|^2 | U_{\mu 3}|^2 \sin^2 \frac{\Delta m^2_{13} L}{4E} - 4 | U_{\mu 2}|^2 | U_{\mu 3}|^2 \sin^2 \frac{\Delta m^2_{23} L}{4E} ~,
\end{equation}
where the usual notation is used for the mass squared 
differences $\Delta m^2_{ij}$ and for the elements of the leptonic mixing matrix $U$.
In the following we take $\theta_{13}=0$.
The oscillation probabilities depend on whether
$|\Delta m^2_{13}| > |\Delta m^2_{23}|$ (normal hierarchy) or $|\Delta m^2_{13}| < |\Delta m^2_{23}|$ (inverted hierarchy).
Precisely measured disappearance probabilities can distinguish between normal and inverted hierarchies if there is sensitivity to effects driven by both $|\Delta m^2_{13}|$ and  $\Delta m^2_{12}$.
This requires the atmospheric mass squared difference to be measured at different $L/E$ with a precision of better than $|\Delta m^2_{21}| / |\Delta m^2_{31}| \sim 0.026$. In fact, it was pointed out in Ref.~\cite{deGouvea:2005hk} that, for a fixed $L/E$, the disappearance
probabilities for the normal and inverted hierarchies are 
the same if $|\Delta m^2_{13}|$ is substituted with
$-|\Delta m^2_{13}| +  \Delta m^2_{12} + \frac{4 E}{L} \arctan \Big(
\cos {2 \theta_{12} } \tan \frac{  \Delta m^2_{12} L}{4E} \Big)$.
In order to break this degeneracy it is necessary to measure the atmospheric mass squared difference at different energies and at distances for which the oscillations driven by the solar term are non negligible.
In our setup, if we assume a $0\%$ ($2\%$) overall systematic error, we find that the hierarchy can be measured at the $1 \sigma$ level ($1 \sigma$ level) for the $10^{23}$~Kton-decays case, while for the $3 \times 10^{23}$~Kton-decays scenario it can be determined at $4 \sigma$ level ($2 \sigma$ level). Note that the systematic errors play a crucial role. It is in principle possible to reduce the impact of the systematics errors using the ratios of the number of events at the near and far detectors:
\begin{equation}
{\cal R} (E)  = \frac{\frac{N_{\mathrm N} (\nu_\mu) }{N_{\mathrm N} (\bar{\nu}_e) }}{\frac{N_{\mathrm F} (\nu_\mu) }{N_{\mathrm F} (\bar{\nu}_e) }}~,
\end{equation}
where $N_{\mathrm N (F)} (\nu_\mu [\bar{\nu}_e]) $ refer to the number of $\nu_\mu \ [\bar{\nu}_e]$ events in the near (far) detector for a fixed energy $E$. 
Very good energy resolution is required for such cancellations to be effective.
In this case, a low energy Neutrino Factory 
can give important information on the type of hierarchy 
even if $\theta_{13}=0$.

\subsection{Simultaneous fits to $\theta_{13}$ and $\delta$}

Next, we study the extraction of the unknown parameters $\theta_{13}$ and $\delta$, using the \emph{golden channel} ($\nu_e (\bar{\nu}_e) \to \nu_\mu (\bar{\nu}_\mu)$).
We start by considering a neutrino factory scenario with $1 \times 10^{23}$~Kton-decays. We find that, for values of $\theta_{13}>2^\circ$, the sign degeneracy is resolved at the $95\%$~C.L. Note that for $\theta_{13} >4^\circ$ the octant degeneracy has already been resolved using the disappearance data. 

\begin{figure}[h]
\begin{center}
\begin{tabular}{ll}
\includegraphics[width=3in]{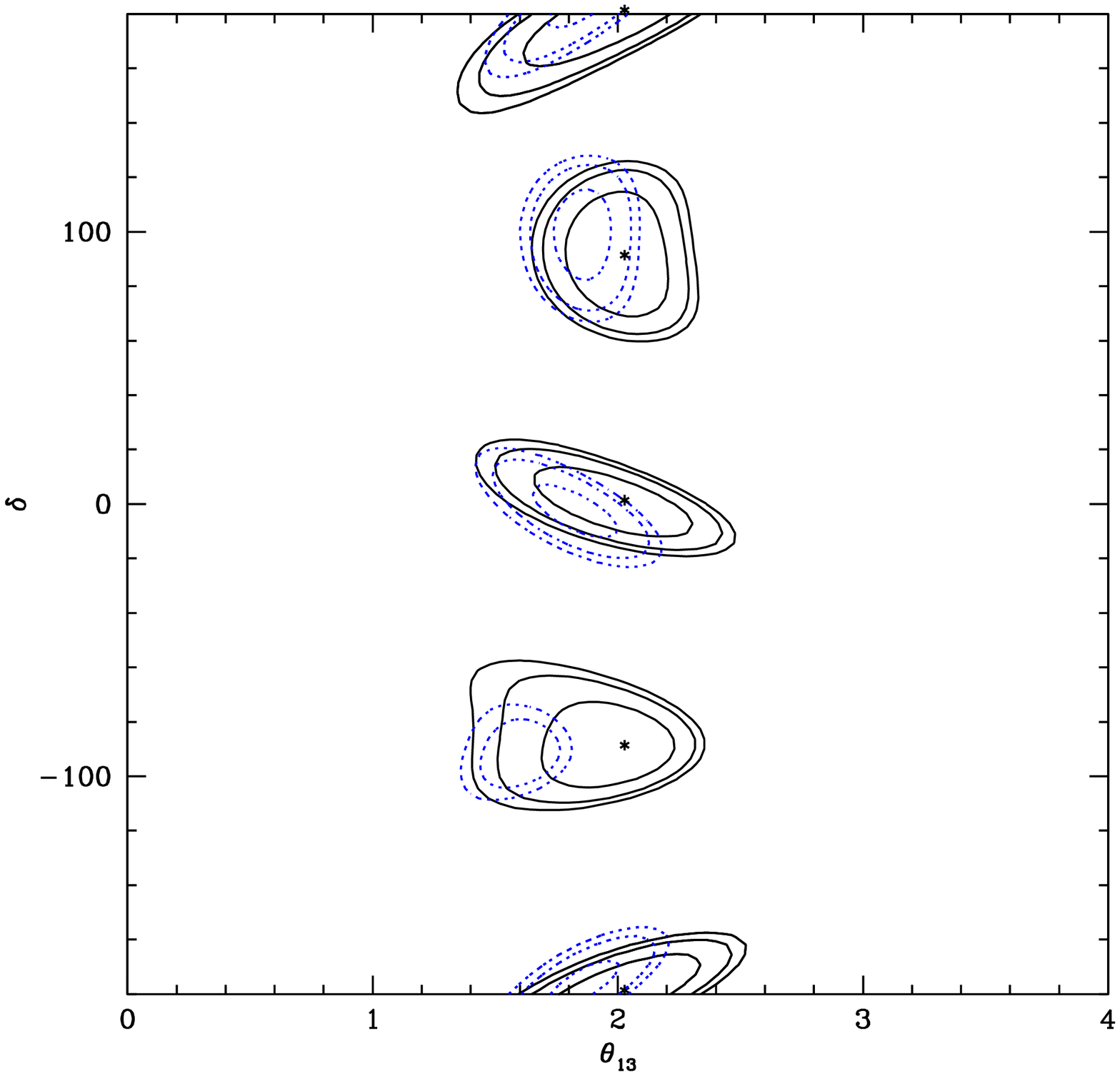}&\hskip 0.cm
\includegraphics[width=3in]{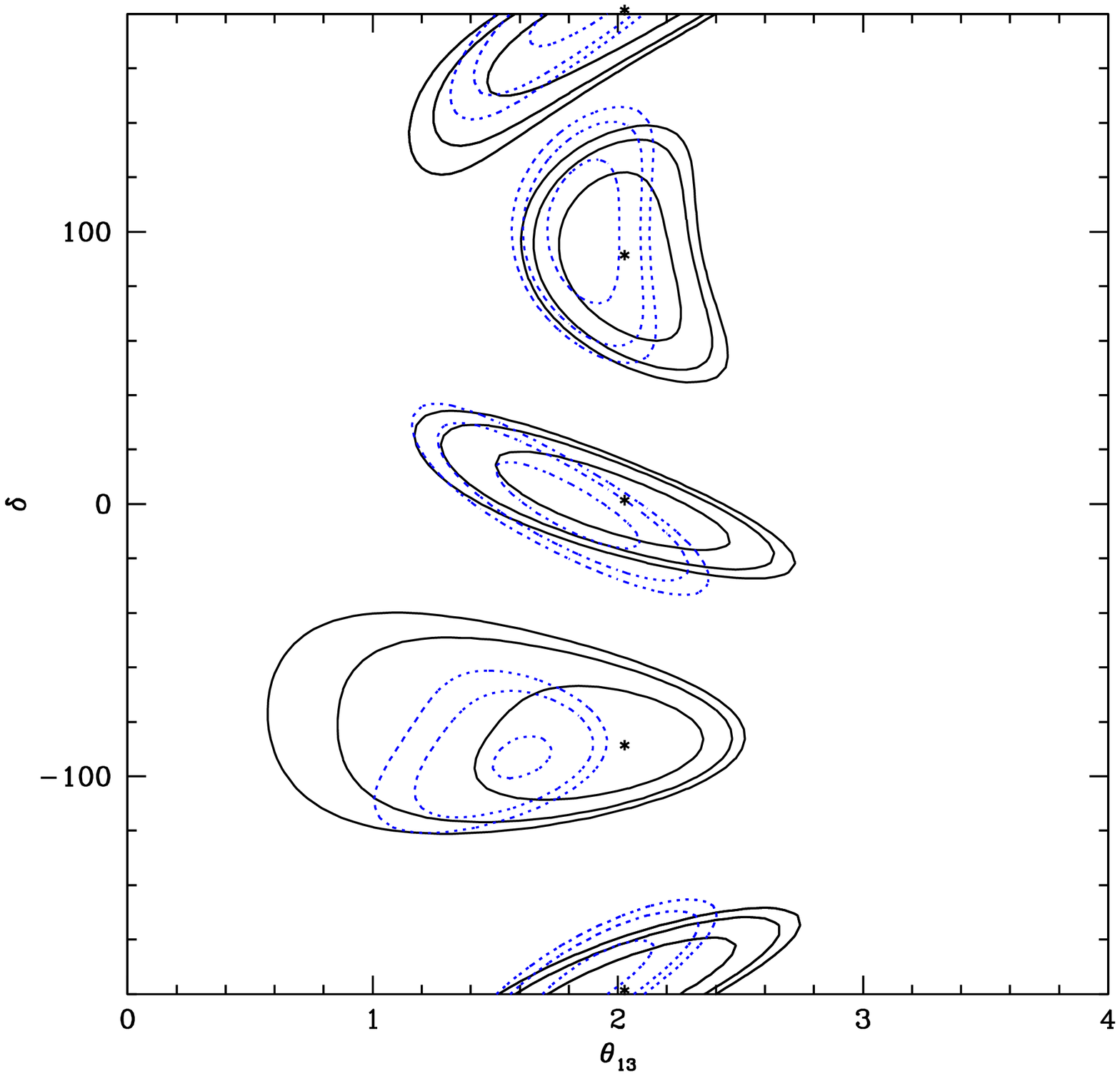}\\
\hskip 2.truecm
{\small}            &
\hskip 2.truecm
{\small}       \\  
\end{tabular}
\end{center}
\caption{\textit{$68\%$, $90\%$ and $95\%$ (2 d.o.f) CL contours resulting from the fits at $L = 1480$ km assuming four central values for $\delta=0^{\circ}$, $90^{\circ}$, $-90^{\circ}$ and $180^{\circ}$ and  $\tetaot=2^\circ$ without backgrounds (left panel) and with a background level of $10^{-3}$ (right panel). The additional $\theta_{23}$ octant solutions are depicted in dotted blue. The statistics considered for both simulations corresponds to $10^{23}$~Kton-decays.}}
\label{fig:fig2}
\end{figure}

Figure~\ref{fig:fig2} shows, for a fit to the simulated data at a baseline $L=1480$ km, the $68\%$, $90\%$ and $95\%$~CL contours in the ($\theta_{13}, \delta$)-plane. Results are shown for background levels set to zero (left panel) and $10^{-3}$ (right panel) for the $10^{23}$~Kton-decays scenario. The four sets of contours correspond to four simulated test points in the ($\theta_{13}, \delta$)-plane, which are depicted by a star. The simulations are for the normal mass hierarchy and $\theta_{23}$ in the first octant ($\sin^2 \theta_{23} = 0.41$ which corresponds to $\theta_{23}=40^\circ$). Our analysis includes the study of the discrete degeneracies. That is, we have fitted the data assuming both the right and wrong hierarchies, and the right and wrong choices for the $\theta_{23}$ octant. If present, the additional solutions associated to the $\theta_{23}$ octant ambiguity are shown as dotted contours.

Notice from Fig.~\ref{fig:fig2} that the sign ambiguity is resolved at the $95\%$ CL in the $10^{23}$~Kton-decays scenario. Additional solutions associated to the wrong choice of the $\theta_{23}$ octant are still present in the $10^{23}$ Kton-decays scenario, but notice that the presence of these additional solutions does not interfere with a measurement of the CP violating phase $\delta$ and $\theta_{13}$, since the locations of the  fake solutions in the ($\theta_{13}$, $\delta$) plane, are almost the same as the correct locations.

The effect of the background can be easily understood in terms of the statistics presented in Tables~\ref{tab:tab2} and \ref{tab:tab3}. For small values of $\theta_{13}$, the addition of the background has a larger impact for $\delta \sim -90^\circ$, since for that value of the CP phase the statistics are dominated by the antineutrino channel, which suffers from a larger background (from $\nu_\mu$'s) than the neutrino channel (from $\bar{\nu}_\mu$'s). For a background level smaller than $\sim 10^{-4}$, the results are indistinguishable from the zero background case. 

\begin{figure}[h]

\begin{center}

\begin{tabular}{ll}

\includegraphics[width=3in]{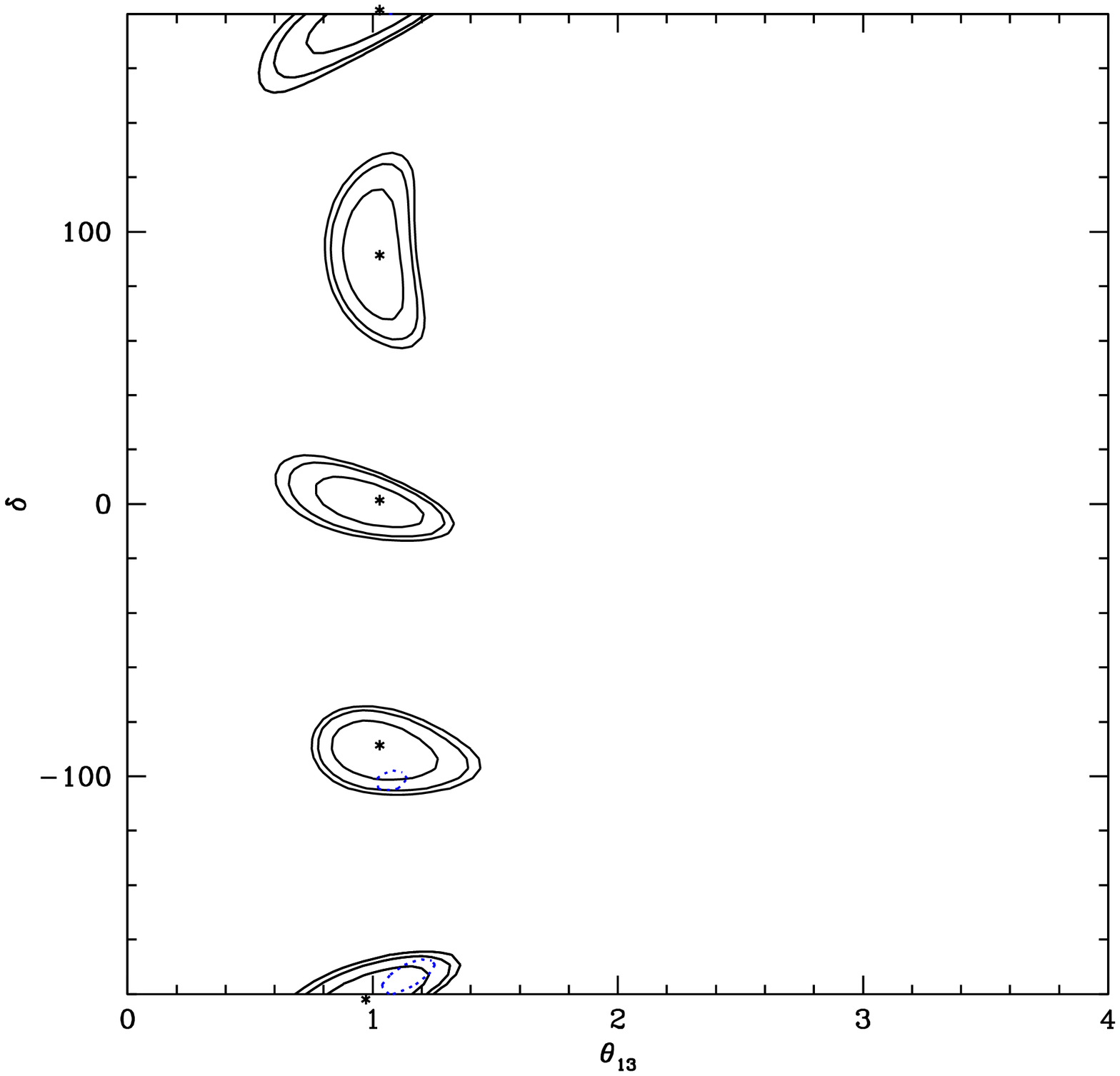}&\hskip 0.cm
\includegraphics[width=3in]{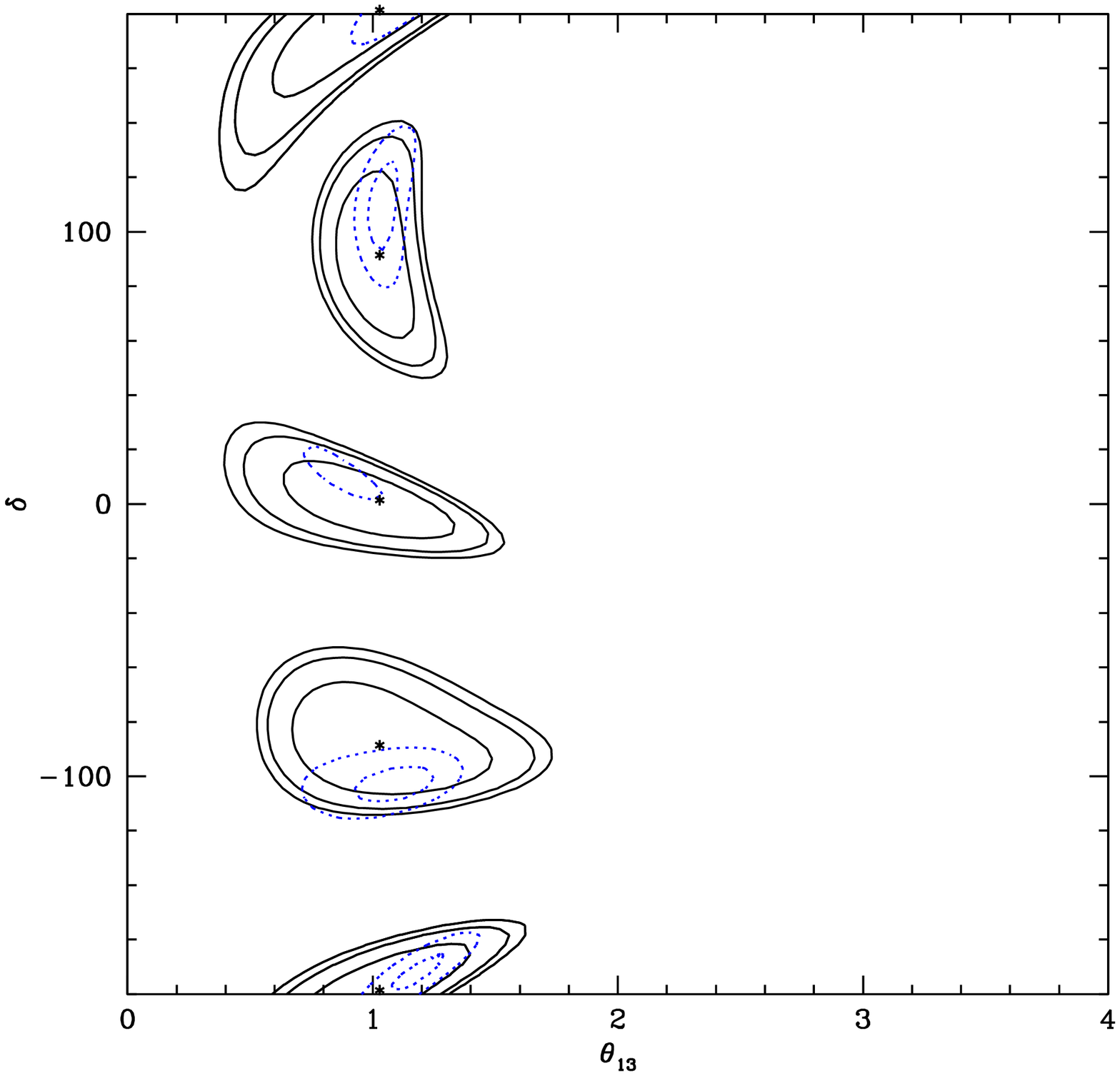}\\
\hskip 2.truecm
{\small}            &
\hskip 2.truecm
{\small}       \\  
\end{tabular}
\end{center}
\caption{\textit{$68\%$, $90\%$ and $95\%$ (2 d.o.f) CL contours resulting from the fits at $L = 1480$ km assuming four central values for $\delta=0^{\circ}$, $90^{\circ}$, $-90^{\circ}$ and $180^{\circ}$ and  $\tetaot=1^\circ$ without backgrounds (left panel) and with a background level of $10^{-3}$ (right panel). The statistics considered for both simulations corresponds to $3\times 10^{23}$~Kton-decays.}}
\label{fig:fig1}

\end{figure}

We illustrate the corresponding results for the improved scenario of $3\times 10^{23}$~Kton-decays in Fig.~\ref{fig:fig1}. Note that the higher statistics allow us to consider a smaller value for $\tetaot=1^\circ$. The additional solutions arising from the wrong choice for the neutrino mass hierarchy or $\theta_{23}$ octant are not present at the $95\%$ CL. Furthermore, the addition of a background level of $10^{-3}$ does not significantly affect the resolution of the degeneracies, and has only an impact on the CP violation measurement.

The performance of the \emph{low energy neutrino factory} in the two high statistics scenarios explored here is unique. The sign($\Delta m_{31}^2$) can be determined at the $95\%$~CL in the $10^{23}$~Kton-decays ($3 \times 10^{23}$~Kton-decays) scenario if $\theta_{13}>2^\circ$ ($>1^\circ$) for all values of the CP phase $\delta$. The $\theta_{23}$-octant ambiguity can be removed at the $95\%$ CL down to roughly $\theta_{13}>0.5-1.0^\circ$  for the representative  choice of $\sin^2 \theta_{23}=0.41$, independently of the value of $\delta$, except for some intermediate values of $\theta_{13}\sim 2^\circ$, for which the $\theta_{23}$ degeneracy is still present for some values of the CP violating phase $\delta$. Resolving the $\theta_{23}$-octant degeneracy therefore is easier for small values of $\theta_{13}<2^\circ$. This is due to the fact that, as explored in Ref.~\cite{GeerMenaPascoli}, the $\theta_{23}$-octant degeneracy is resolved using the information from the low energy bins, which are sensitive to the solar term. For the setup described in this paper, the solar term starts to be important if $\theta_{13}<2^\circ$. However, notice that the presence of the $\theta_{23}$ octant ambiguity at $\theta_{13}\sim 2^\circ$ will not interfere with the extraction of $\theta_{13}$ and $\delta$, since the locations of the degenerate (fake) solutions almost coincide with the positions of ``true'', nature solutions.


\begin{figure}[h]
\begin{center}
\includegraphics[width=3.5in]{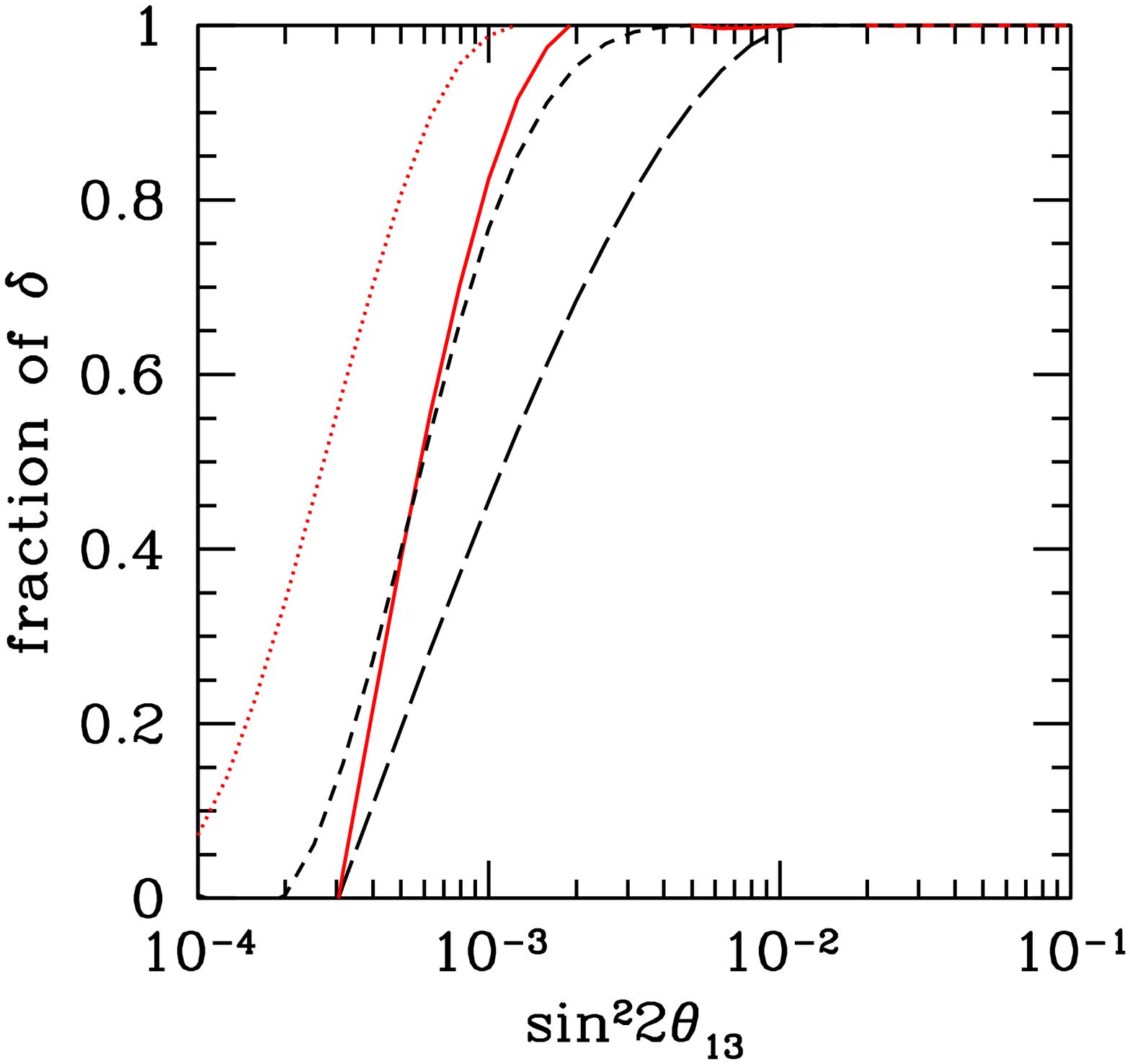}
\end{center}
\caption[]{\textit{$95\%$ CL (1 d.o.f) hierarchy resolution assuming that the far detector is located at a distance of $1480$ km at the Henderson mine. The solid (dotted) red curves depict the results assuming  $1\times 10^{23}$~Kton-decays ($3\times 10^{23}$~Kton-decays) without backgrounds.  The long-dashed (short-dashed) black curves depict the results assuming  $1\times 10^{23}$~Kton-decays ($3\times 10^{23}$~Kton-decays) with a background level of $10^{-3}$. }}
\label{fig:hier}
\end{figure}

\begin{figure}[h]
\begin{center}
\includegraphics[width=3.5in]{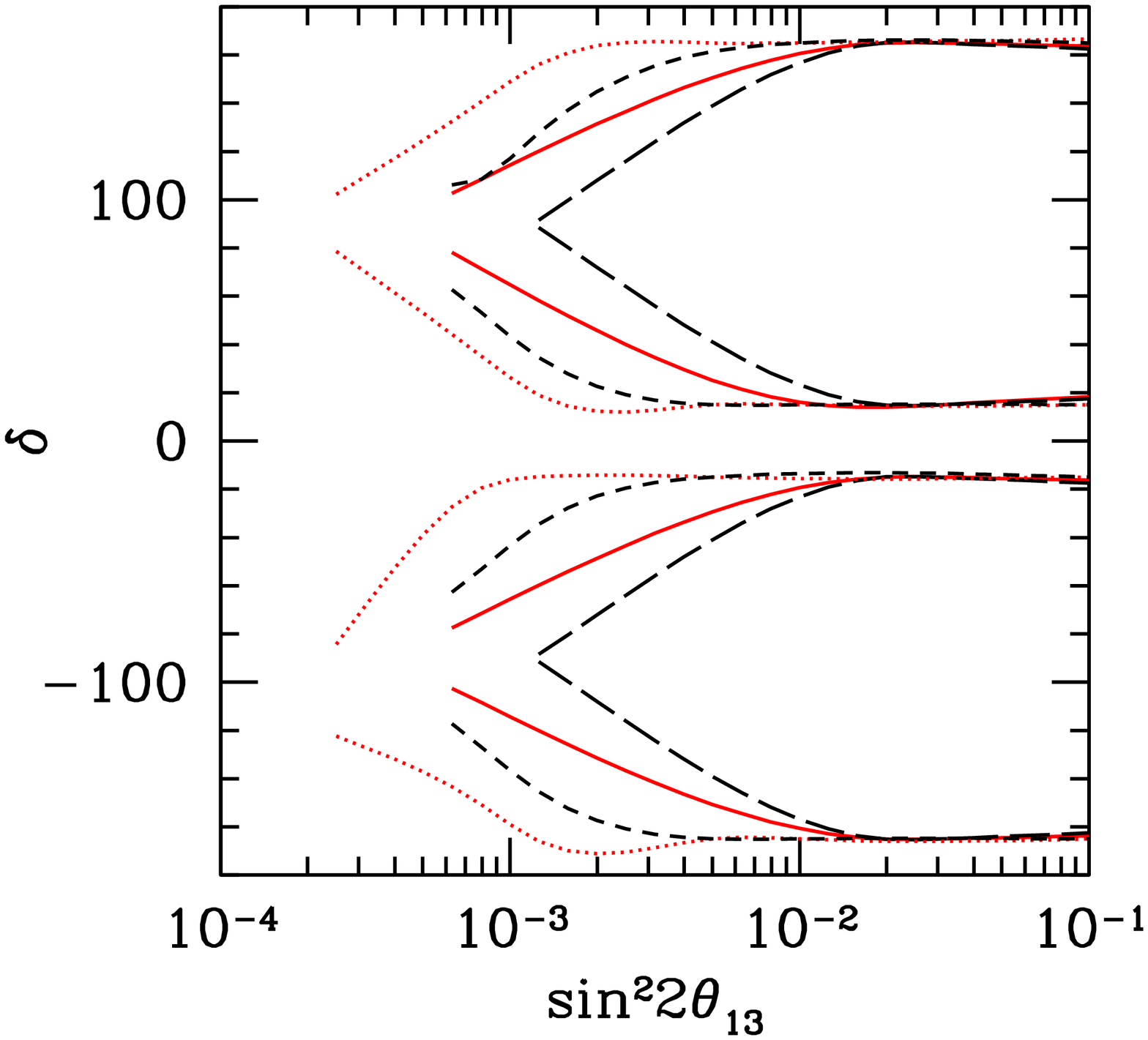}
\end{center}
\caption[]{\textit{$95\%$ CL (1 d.o.f) CP Violation  extraction assuming that the far detector is located at a distance of $1480$. The solid (dotted) red curves depict the results assuming  $1\times 10^{23}$~Kton-decays ($3\times 10^{23}$~Kton-decays) without backgrounds. The long-dashed (short-dashed) black curves depict the results assuming  $1\times 10^{23}$~Kton-decays ($3\times 10^{23}$~Kton-decays) with a background level of $10^{-3}$.}}
\label{fig:cp}
\end{figure}

In Figs.~\ref{fig:hier} and \ref{fig:cp} we summarize, for the $10^{23}$ and $3\times 10^{23}$~Kton decays scenarios, the physics reach for a TASD detector located 1480 km from a low energy neutrino factory. The analysis takes into account the impact of both the intrinsic and discrete degeneracies. Figure~\ref{fig:hier} shows the region in the ($\sin^2 2 \theta_{13}$, ``fraction of $\delta$'') plane for which the mass hierarchy can be resolved at the $95\%$ CL (1 d.o.f). Contours are shown for zero background, and for when a background level of $10^{-3}$ is included in the analysis. Note that, with a background level of $\simeq 10^{-3}$, the hierarchy can be determined in both scenarios if $\sin^2 2 \theta_{13}>$ few $10^{-3}$ (i.e. $\theta_{13}> 2-3^\circ$) for all values of the CP violating phase $\delta$. For a background level smaller than $\sim 10^{-4}$, the results are indistinguishable from the zero background case.

Figure~\ref{fig:cp} shows the region in the ($\sin^2 2 \theta_{13}$, $\delta$) plane for which a given (non-zero) CP violating value of the CP-phase $\delta$ can be distinguished at the $95\%$ CL (1 d.o.f) from the CP conserving case, i.e. $\delta =0, \pm 180^\circ$. The results are given for the two statistics scenarios studied here.  Note that, even in the presence of a $10^{-3}$ background level, the CP violating phase $\delta$ could be measured with a $95\%$ CL precision of better than $20^\circ$ in the $10^{23}$~Kton-decays ($3\times 10^{23}$~kton-deacys) luminosity scenario if $\sin^2 2 \theta_{13}>0.01$ ($\sin^2 2 \theta_{13}>0.002$).

\section{Summary and Conclusions}

We have studied the physics reach of a \emph{low energy neutrino factory}, first presented in Ref.~\cite{GeerMenaPascoli}, in which the stored muons have an energy of $4.12$~GeV. The simulated detector performance is based upon a 
magnetized Totally Active Scintillator Detector. Our simulations suggest this detector will have a threshold for measuring muon neutrino CC interactions of about 500 MeV and an energy independent efficiency of about 73\% above threshold. We have assumed a conservative energy resolution of 30\% for the detector.

In our analysis, we consider the representative baseline of $1480$ Km, divide the simulated observed neutrino event spectrum into 9 energy bins above the 500~MeV threshold, and exploit both the disappearance ($\nu_\mu \to \nu_\mu$)  and the \emph{golden} ($\nu_e \to \nu_\mu$) channels by measuring CC events tagged by ``right-sign'' and ``wrong-sign'' muons.  The results can be easily generalized to other baselines in the 1200--1500~km range. We have investigated the dependence of the physics sensitivity on statistics by considering a high statistics scenario corresponding to  $1 \times 10^{23}$~Kton-decays for each muon sign, and a more aggressive scenario corresponding to $3 \times 10^{23}$~Kton-decays for each muon sign. We have also explored the impact of backgrounds to the wrong-sign muon signal by considering background levels of zero and $10^{-3}$.

We find that, based only on the disappearance channel, maximal atmospheric neutrino mixing can be excluded at $95\%$ CL if $\sin^2 \theta_{23}<0.44$ ($\theta_{23}<41.5^\circ$). The atmospheric mass difference could be measured with a precision of $1\%-2\%$, opening the possibility of determining the neutrino mass hierarchy even if $\theta_{13}=0$, provided systematic uncertainties can be controlled. Neglecting systematic uncertainties, the mass  hierarchy could be determined at the $1 \sigma$ level ($4 \sigma$ level) in the $1 \times 10^{23}$~Kton-decays ($3 \times 10^{23}$~Kton-decays) statistics scenario.

The rich oscillation pattern of the $\nu_e\to \nu_\mu$ ($\bar\nu_e \to \bar\nu_\mu$) appearance channels at energies between $0.5$ and $4$ GeV for baselines $\mathcal{O}(1000)$~km facilitates an elimination of the degenerate solutions.  If the atmospheric mixing angle is not maximal, for the representative choice of $\sin^2 \theta_{23}=0.4$, the octant in which $\theta_{23}$ lies could be extracted at the $95\%$ CL in both scenarios if $\theta_{13}> 0.5-1^\circ$, for all values of the CP violating phase $\delta$, except for some intermediate values of the mixing angle $\theta_{13}$ in which the fake solutions's location coincides with the true's solution position and therefore the presence of these fake solutions does not interfere with the extraction of $\delta_{CP}$ and $\theta_{13}$.

In the $10^{23}$ kton-decays scenario, if the background level is $\sim 10^{-3}$ ($10^{-4}$), the neutrino mass hierarchy could be determined at the $95\%$ CL, and the CP violating phase $\delta$ could be measured with a $95\%$ CL precision of better than $20^\circ$, if $\sin^2 2 \theta_{13}>0.01$ ($\sim 0.006$). With a factor of three improvement in the former statistics, the numbers quoted above are $\sin^2 2 \theta_{13}= 0.005$ and $\sin^2 2 \theta_{13}=0.002$, for background levels of $10^{-3}$ and $10^{-4}$, respectively. In our analysis we have included a $2\%$ systematic error on all measured event rates.

In summary, the low statistics low energy Neutrino Factory scenario we have described, with a background level of $10^{-3}$, for both large and very small values of $\theta_{13}$ would be able to eliminate ambiguous solutions, determine $\theta_{13}$, the mass hierarchy, and search for CP violation. Higher statistics and lower backgrounds would further improve the sensitivity, and may enable the mass hierarchy to be determined even if $\theta_{13}=0$.

\vspace{1cm}
\section*{Acknowledgments} 
This work was supported in part by the European Programme ``The Quest for
Unification''  contract MRTN-CT-2004-503369, and by the Fermi National Accelerator Laboratory, which is operated by the Fermi Research Association, under contract No. DE-AC02-76CH03000 with the U.S. Department of Energy. SP acknowledges the support of CARE,
contract number RII3-CT-2003-506395.
OM and SP would like to thank the Theoretical Physics Department at Fermilab for hospitality and support. 
\section*{Appendix} 
\begin{figure}[h]
\begin{center}
\includegraphics[width=6in]{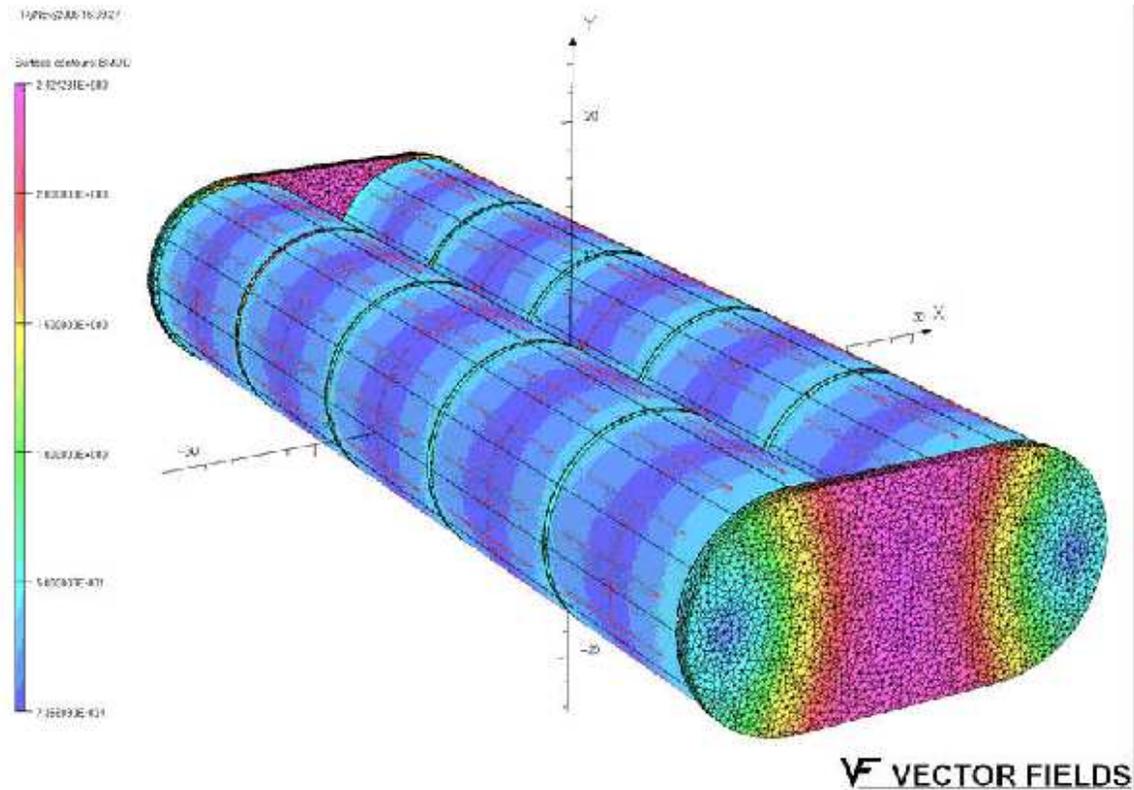}
\end{center}
\caption[]{\textit{Simulation results for magnetic cavern design.}}
\label{fig:mag_cav}
\end{figure}

All detector concepts for the Neutrino Factory (NF) require a magnetic field in order to determine the sign of muon (or possibly the electron) produced in the neutrino interaction. For the baseline detector, this is done with magnetized iron. Technically this is very straightforward, although for the $100$~Kton baseline detector does present challenges because of its size. The cost of this magnetic solution is felt to be manageable. Magnetic solutions for the TASD become much more problematic.  The solution that we propose is to use the Superconducting Transmission Line developed for the VLHC~\cite{VLHC}, Fig.~\ref{fig:STL}, as windings for very large solenoids that form a magnetic cavern (see Fig.~\ref{fig:mag_cav}) for the detector.  The Superconducting Transmission Line (STL) consists of a superconducting cable inside a cryo-pipe cooled by supercritical liquid helium at $4.5-6.0$~ K placed inside a co-axial cryostat. It consists of a perforated Invar tube, a copper stabilized superconducting cable, an Invar helium pipe, the cold pipe support system, a thermal shield covered by multilayer superinsulation, and the vacuum shell. One of the possible STL designs developed for the VLHC is shown in Fig.~\ref{fig:mag_cav} within the main text.  The STL is designed to carry a current of $100$~kA at $6.5$~K in a magnetic field up to $1$~T. This provides a $50\%$ current margin with respect to the required current in order to reach a field of $0.5$~T.  This operating margin can compensate for temperature variations, mechanical or other perturbations in the system.

\begin{figure}[h]
\begin{center}
\includegraphics[width=6in]{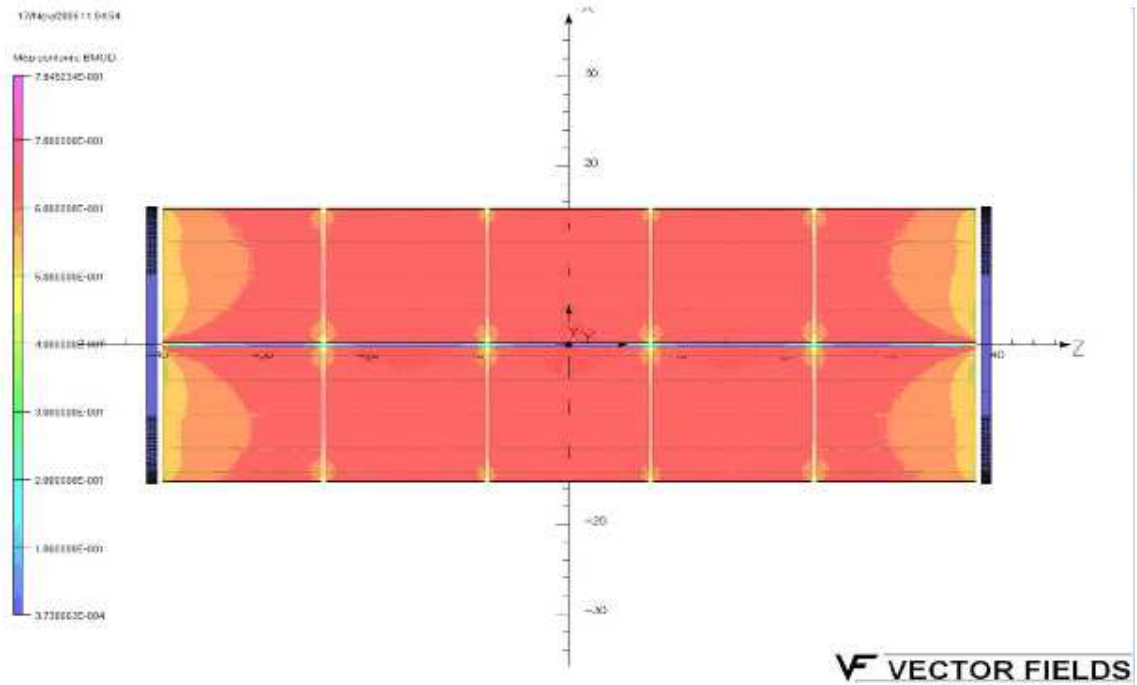}
\end{center}
\caption[]{\textit{STL Solenoid Magnetic Cavern Field Uniformity in XZ plane.}}
\label{fig:stl_sol}
\end{figure}

\begin{figure}[h]
\begin{center}
\includegraphics[width=6in]{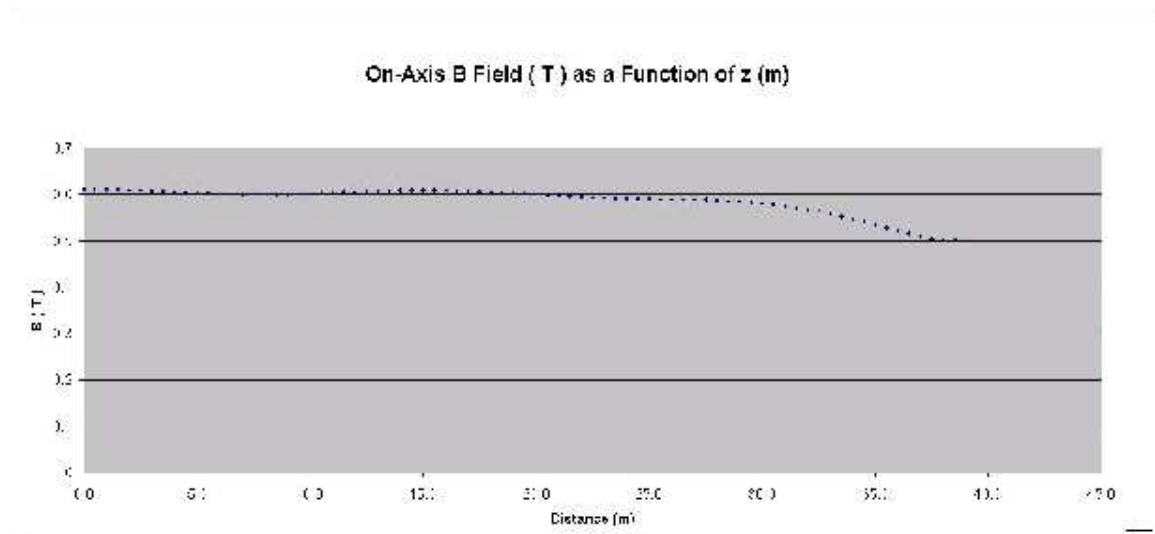}
\end{center}
\caption[]{\textit{On-axis $B$ field in $T$ as a function of position along the $z$ axis (m).}}
\label{fig:app3}
\end{figure}

The solenoid windings now consist of this superconducting cable which is confined in its own cryostat.  Each solenoid consists of $150$ turns and requires $\sim 7500$~ m of cable. There is no large vacuum vessel and access to the detectors can be made through the winding support cylinder since the STL does not need to be close-packed in order to reach an acceptable field. We have performed a simulation of the Magnetic Cavern concept using STL solenoids and the results are shown in Fig.~\ref{fig:stl_sol}.  With the iron end-walls ($1$~m thick), the average field in the XZ plane is approximately $0.58$~T at an excitation current of $50$~kA. This figure shows the field uniformity in the X-Z plane which is better than $\pm 2\% $ throughout the majority of the volume with approximately $20\%$ variations near the end-irons. Figure~\ref{fig:app3} shows the on-axis $B$ field as a function of position along the $z$ axis (in meters).

We have not yet been able to do a detailed costing of the magnetic cavern.  The STL costs can be estimated quite accurately ($30\%$) from the VLHC work and current SC cable costs and are believed to be $\$50$M.  The total magnetic cavern cost is estimated to be less than $\$150$M.  This is to be compared to a fully loaded cost savings of the Low-Energy Neutrino Factory (compared to the $50$~GeV design) as indicated in Ref.~\cite{ISS-Detector Report}.


\begin{thebibliography}{99}
\bibitem{SKatm}
  Y.~Ashie {\it et al.}  [Super-Kamiokande Collaboration],
  Phys.\ Rev.\ D {\bf 71}, 112005 (2005).
\bibitem{sol}
  B.~T.~Cleveland {\it et al.},
  Astrophys.\ J.\  {\bf 496}, 505 (1998);
  Y.~Fukuda {\em et al.} [Kamiokande Collaboration],
  Phys.\ Rev.\ Lett.\  {\bf 77}, 1683 (1996);
  J.~N.~Abdurashitov {\it et al.}  [SAGE Collaboration],
  J.\ Exp.\ Theor.\ Phys.\  {\bf 95}, 181 (2002);
  W.~Hampel {\it et al.}  [GALLEX Collaboration],
  Phys.\ Lett.\ B {\bf 447}, 127 (1999);
  T.~A.~Kirsten  [GNO Collaboration],
  Nucl.\ Phys.\ Proc.\ Suppl.\  {\bf 118}, 33 (2003).
\bibitem{SKsolar}
  S.~Fukuda {\it et al.}  [Super-Kamiokande Collaboration],
  Phys.\ Lett.\ B {\bf 539}, 179 (2002).
\bibitem{SNO1}
  Q.~R.~Ahmad {\it et al.}  [SNO Collaboration],
  Phys.\ Rev.\ Lett.\  {\bf 87}, 071301 (2001).
\bibitem{SNO2}
  Q.~R.~Ahmad {\it et al.} [SNO Collaboration],
  Phys.\ Rev.\ Lett.\  {\bf 89}, 011301 (2002)
  and {\it ibid.} {\bf 89}, 011302 (2002).
\bibitem{SNO3}
  S.~N.~Ahmed {\it et al.} [SNO Collaboration],
  Phys.\ Rev.\ Lett.\  {\bf 92}, 181301 (2004).
\bibitem{SNOsalt}
  B.~Aharmim {\it et al.}  [SNO Collaboration],
  Phys.\ Rev.\ C {\bf 72}, 055502 (2005).
\bibitem{KamLAND}
  K.~Eguchi {\it et al.}  [KamLAND Collaboration],
  Phys.\ Rev.\ Lett.\  {\bf 90}, 021802 (2003).
\bibitem{K2K}
  M.~H.~Ahn {\it et al.}  [K2K Collaboration],
  Phys.\ Rev.\  D {\bf 74}, 072003 (2006).
\bibitem{MINOS}
  D.~G.~Michael {\it et al.}  [MINOS Collaboration],
  Phys.\ Rev.\ Lett.\  {\bf 97}, 191801 (2006).
\bibitem{thomas}
  T.~Schwetz,
  Phys.\ Scripta {\bf T127}, 1 (2006).
\bibitem{newfit}
  G.~L.~Fogli {\it et al.},
  Phys.\ Rev.\  D {\bf 75}, 053001 (2007).
\bibitem{concha}
  M.~C.~Gonzalez-Garcia and M.~Maltoni,
  arXiv:0704.1800 [hep-ph].
\bibitem{geer}
  S.~Geer,
  Phys.\ Rev.\ D {\bf 57}, 6989 (1998)
  [Erratum-ibid.\ D {\bf 59}, 039903 (1999)].
  
\bibitem{nf1}
  A.~De Rujula, M.~B.~Gavela and P.~Hernandez,
  Nucl.\ Phys.\ B {\bf 547}, 21 (1999).
\bibitem{nf2}
  V.~D.~Barger, S.~Geer and K.~Whisnant,
  Phys.\ Rev.\ D {\bf 61}, 053004 (2000);
  A.~Donini {\it et al}, 
  Nucl.\ Phys.\ B {\bf 574}, 23 (2000);
  V.~D.~Barger {\it et al}, 
  Phys.\ Rev.\ D {\bf 62}, 073002 (2000).
  
\bibitem{nf4}
  V.~D.~Barger {\it et al}, 
  Phys.\ Rev.\ D {\bf 62}, 013004 (2000).
  
\bibitem{nf5}
  A.~Cervera, {\it et al}
  Nucl.\ Phys.\ B {\bf 579}, 17 (2000)
  [Erratum-ibid.\ B {\bf 593}, 731 (2001)];
\bibitem{nf6}
  M.~Freund, P.~Huber and M.~Lindner,
  Nucl.\ Phys.\ B {\bf 585}, 105 (2000);
  V.~D.~Barger {\it et al}, 
  Phys.\ Lett.\ B {\bf 485}, 379 (2000);
  J.~Burguet-Castell {\it et al}, 
  Nucl.\ Phys.\ B {\bf 608}, 301 (2001);
  M.~Freund, P.~Huber and M.~Lindner,
  Nucl.\ Phys.\ B {\bf 615}, 331 (2001).
  
\bibitem{silver}
  A.~Donini, D.~Meloni and P.~Migliozzi,
  Nucl.\ Phys.\ B {\bf 646}, 321 (2002);
  
  D.~Autiero {\it et al.},
  Eur.\ Phys.\ J.\ C {\bf 33}, 243 (2004).
  
\bibitem{study1-physics}
  C.~Albright {\it et al.},
  arXiv:hep-ex/0008064.


\bibitem{nf8}
  A.~Blondel {\it et al.},
  Nucl.\ Instrum.\ Meth.\ A {\bf 451}, 102 (2000);

  M.~Apollonio {\it et al.},
  arXiv:hep-ph/0210192;
\bibitem{study2}
C.~Albright {\it et al.}  [Neutrino Factory/Muon Collider Collaboration],
arXiv:physics/0411123.
\bibitem{yo}
  O.~Mena,
  Mod.\ Phys.\ Lett.\ A {\bf 20}, 1 (2005).
\bibitem{nf9}
  P.~Huber, M.~Lindner, M.~Rolinec and W.~Winter,
  Phys.\ Rev.\  D {\bf 74}, 073003 (2006).
\bibitem{GeerMenaPascoli}
  S.~Geer, O.~Mena and S.~Pascoli,
  Phys.\ Rev.\  D {\bf 75}, 093001 (2007).
\bibitem{hubernew}
P.~Huber and W.~Winter,
  arXiv:0706.2862 [hep-ph].
\bibitem{FL96} 
G.~L.~Fogli and E.~Lisi,
  Phys.\ Rev.\ D {\bf 54}, 3667 (1996).
\bibitem{MN01}
  H.~Minakata and H.~Nunokawa,
  JHEP {\bf 0110}, 001 (2001).
\bibitem{BMWdeg} 
V.~D.~Barger, S.~Geer, R.~Raja and K.~Whisnant,
  Phys.\ Rev.\ D {\bf 63}, 113011 (2001).
\bibitem{deg}
  T.~Kajita, H.~Minakata and H.~Nunokawa,
  Phys.\ Lett.\ B {\bf 528}, 245 (2002);
  H.~Minakata, H.~Nunokawa and S.~J.~Parke,
  Phys.\ Rev.\ D {\bf 66}, 093012 (2002);
  P.~Huber, M.~Lindner and W.~Winter,
  Nucl.\ Phys.\ B {\bf 645}, 3 (2002);
  A.~Donini, D.~Meloni and S.~Rigolin,
  JHEP {\bf 0406}, 011 (2004);
  M.~Aoki, K.~Hagiwara and N.~Okamura,
  Phys.\ Lett.\ B {\bf 606}, 371 (2005);
  O.~Yasuda,
  New J.\ Phys.\  {\bf 6}, 83 (2004);
  O.~Mena and S.~J.~Parke,
  Phys.\ Rev.\ D {\bf 72}, 053003 (2005).
\bibitem{MN97}
  H.~Minakata and H.~Nunokawa,
  Phys.\ Lett.\ B {\bf 413}, 369 (1997).
\bibitem{BMW02off}
  V.~Barger, D.~Marfatia and K.~Whisnant,
  Phys.\ Rev.\ D {\bf 66}, 053007 (2002).
\bibitem{SN1}
  O.~Mena Requejo, S.~Palomares-Ruiz and S.~Pascoli,
  Phys.\ Rev.\ D {\bf 72}, 053002 (2005).
\bibitem{twodetect}
  M.~Ishitsuka {\it et al.}, 
  Phys.\ Rev.\ D {\bf 72}, 033003 (2005);
  K.~Hagiwara, N.~Okamura and K.~i.~Senda,
  Phys.\ Lett.\ B {\bf 637}, 266 (2006).
\bibitem{SN2}
O.~Mena, S.~Palomares-Ruiz and S.~Pascoli,
  Phys.\ Rev.\ D {\bf 73}, 073007 (2006).
\bibitem{T2kk}
  T.~Kajita {\it et al},
  arXiv:hep-ph/0609286.
\bibitem{otherexp1}
  J.~Burguet-Castell {\it et al.},
  Nucl.\ Phys.\ B {\bf 646}, 301 (2002);
\bibitem{HLW02}
  P.~Huber, M.~Lindner and W.~Winter,
  Nucl.\ Phys.\ B {\bf 654}, 3 (2003).
\bibitem{MNP03}
  H.~Minakata, H.~Nunokawa and S.~J.~Parke,
  Phys.\ Rev.\ D {\bf 68}, 013010 (2003).
\bibitem{BMW02}
  V.~Barger, D.~Marfatia and K.~Whisnant,
  Phys.\ Lett.\ B {\bf 560}, 75 (2003).
\bibitem{otherexp}
  K.~Whisnant, J.~M.~Yang and B.~L.~Young,
  Phys.\ Rev.\ D {\bf 67}, 013004 (2003);
  P.~Huber {\it et al.},
  Nucl.\ Phys.\ B {\bf 665}, 487 (2003);
  P.~Huber {\it et al.},
  Phys.\ Rev.\ D {\bf 70}, 073014 (2004);
  A.~Donini, E.~Fern\'andez-Mart\'{\i}nez and S.~Rigolin,
  Phys.\ Lett.\ B {\bf 621}, 276 (2005).
\bibitem{mp2}
  O.~Mena and S.~J.~Parke,
  Phys.\ Rev.\ D {\bf 70}, 093011 (2004).
\bibitem{HMS05}
  P.~Huber, M.~Maltoni and T.~Schwetz,
  Phys.\ Rev.\ D {\bf 71}, 053006 (2005).
\bibitem{Choubey}
S.~Choubey and P.~Roy,
  Phys.\ Rev.\  D {\bf 73}, 013006 (2006).
\bibitem{huber2}
  A.~Blondel {\it et al.},
  Acta Phys.\ Polon.\  B {\bf 37}, 2077 (2006).
A.~Blondel 
  hep-ph/0606111.
\bibitem{lastmine}
  O.~Mena, H.~Nunokawa and S.~J.~Parke,
  Phys.\ Rev.\  D {\bf 75}, 033002 (2007).
  O.~Mena,
  hep-ph/0609031.
  
\bibitem{ISS-Detector Report}
The International Scoping Study for a Neutrino Factory – RAL-TR-2007-24
\bibitem{Nova}
D.~Ayres {\it et al.},
hep-ex/0503053.

\bibitem{Numi}
  J.~Hylen {\it et al.}, FERMILAB-TM-2018, (1997).
  
\bibitem{minerva_www}
  K.~S.~McFarland,
  Eur.\ Phys.\ J.\  A {\bf 24S2}, 187 (2005).
\bibitem{D0}
P.~Baringer {\it et al.}  [D0 Collaboration],
  Nucl.\ Instrum.\ Meth.\  A {\bf 469}, 295 (2001).
\bibitem{VLHC}
Ambrosio et. al, 
Fermilab-TM-2149, June, 2001.
\bibitem{Nuance}
D.~Casper,
  Nucl.\ Phys.\ Proc.\ Suppl.\  {\bf 112}, 161 (2002).
\bibitem{recpack}
  A.~Cervera-Villanueva, J.~J.~Gomez-Cadenas and J.~A.~Hernando,
  Nucl.\ Instrum.\ Meth.\  A {\bf 534}, 180 (2004). 
\bibitem{dis}
  A.~Bueno, M.~Campanelli and A.~Rubbia,
  Nucl.\ Phys.\ B {\bf 589} 577 (2000). 
\bibitem{Stef}
  A.~Donini {\it et al}
  Nucl.\ Phys.\ B {\bf 743}, 41 (2006).


\bibitem{deGouvea:2005hk}
  A.~de Gouvea, J.~Jenkins and B.~Kayser,
  Phys.\ Rev.\  D {\bf 71}, 113009 (2005)
  [arXiv:hep-ph/0503079].

\bibitem{deGouvea:2005mi}
  A.~de Gouvea and W.~Winter,
  Phys.\ Rev.\  D {\bf 73}, 033003 (2006)
  [arXiv:hep-ph/0509359].

\bibitem{Minakata:2006gq}
  H.~Minakata, H.~Nunokawa, S.~J.~Parke and R.~Zukanovich Funchal,
  Phys.\ Rev.\  D {\bf 74}, 053008 (2006)
  [arXiv:hep-ph/0607284].
\end{thebibliography}
\end{document}